\documentclass[12pt,preprint]{aastex}

\slugcomment{Accepted for publication in ApJ}

\shorttitle{High Energy Cosmic-Ray Electrons Observed with ECC}
\shortauthors{T.Kobayashi et al.}

\begin{document}


\title{Observations of High Energy Cosmic-Ray Electrons 
from 30~GeV to 3~TeV with Emulsion Chambers}


\author{T.Kobayashi}
\affil{Department of Physics and Mathematics, Aoyama Gakuin University, 
Sagamihara 252-5258, Japan}
\email{tadasik-112850@jasper.dti.ne.jp}

\author{Y.Komori}
\affil{Kanagawa University of Human Services, Yokosuka 238-0013, Japan}
\email{komori-y@kuhs.ac.jp}

\author{K.Yoshida, K.Yanagisawa}
\affil{College of Systems Engineering and Science, 
Shibaura Institute of Technology, Saitama 337-8570, Japan}
\email{yoshida@shibaura-it.ac.jp}

\author{J.Nishimura, T.Yamagami, Y.Saito}
\affil{The Institute of Space and Astronautical Science, 
Japan Aerospace Exploration Agency, 
Sagamihara 229-8510, Japan}
\email{nisimura@icrr.u-tokyo.ac.jp}

\author{N.Tateyama}
\affil{Faculty of Engineering, Kanagawa University, Yokohama 221-8686, Japan}
\email{tateyama@n.kanagawa-u.ac.jp}

\author{T.Yuda}
\affil{Institute for Cosmic Ray Research, University of Tokyo, Kashiwa 277-8582, Japan}
\email{yuda@icrr.u-tokyo.ac.jp}

\and

\author{R.J.Wilkes}
\affil{Department of Physics, University of Washington, Seatle, USA}
\email{wilkes@u.washington.edu}




\begin{abstract}
We have performed a series of cosmic-ray electron observations using 
the balloon-borne emulsion chambers since 1968.
While we previously reported the results from subsets of the exposures, 
the final results of the total exposures up to 2001 are presented here. 
Our successive experiments 
have yielded the total exposure of 8.19 m$^2$~sr day 
at the altitudes of $4.0 - 9.4$~g~cm$^{-2}$. 
The performance of the emulsion chambers was examined by 
accelerator beam tests and Monte-Carlo simulations, 
and the on-board calibrations were carried out by using the flight data. 
In this work we present the cosmic-ray electron spectrum 
in the energy range from 30~GeV to 3~TeV 
at the top of the atmosphere, 
which is well represented by 
a power-law function with an index of $-3.28{\pm}0.10$. 
The observed data can be also interpreted in terms of 
diffusive propagation models. 
The evidence of cosmic-ray electrons up to 3~TeV suggests 
the existence of cosmic-ray electron sources 
at distances within $\sim1$~kpc and times within 
${\sim}1{\times}10^{5}$~yr ago. 
\end{abstract}



\keywords{cosmic rays: electrons, origin --- balloons --- supernova remnants}


\section{Introduction}
\label{sec:intro}

Electrons\footnote{
In this paper, 
the term "electrons" is used for the sum of particles regardless 
of charge. 
When we must identify the charge, 
we will use the terms "negative electrons" or "positrons".} 
in cosmic rays have unique features, 
complementary to the cosmic-ray nuclear components, 
because of their low mass and leptonic nature. 
High-energy electrons lose energy by synchrotron radiation 
in the Galactic magnetic field and 
inverse Compton scattering with the interstellar photons in the Galaxy. 
High-energy cosmic-ray electrons cannot propagate far from the sources, 
because the electrons lose rapidly energy with an energy loss rate 
of the square of energy through these radiative processes. 
These processes during the propagations through the Galaxy 
without hadronic interactions 
simplify modeling of the propagation of electrons 
compared with other cosmic-ray components such as nucleons.

Evidence for non-thermal X-ray emission from supernova remnants (SNRs) 
indicate that high-energy electrons in the TeV region 
are accelerated in SNRs \citep[e.g.]{koyama95}. 
These observations strongly suggest that 
cosmic-ray electrons are accelerated in SNRs, 
and that SNRs are the most likely primary sources of cosmic-ray electrons. 
\citet{shen70} first pointed out that the electron spectrum in the TeV region 
depends on the age and distance of a few local sources. 
His proposed concept has been accepted in later calculations of cosmic-ray electrons 
\citep[and references therein]{kobayashi04}. 
\citet{kobayashi04} suggest 
that the energy spectrum of cosmic-ray electrons have 
unique spectral structures in the TeV region 
due to the discrete effect of local sources. 
This means that we can identify cosmic-ray electron sources 
from the electron spectrum in the TeV region. 
In addition, it is discussed 
that some dark matter may produce negative electrons and positrons 
in the energy region of around $100 - 10$~TeV 
via dark matter annihilations or decaying dark matter 
\citep[e.g.]{kamionkowski91, cheng02}. 
In particular, 
in the case of mono-energetic electrons from dark matter, 
although propagation through the Galaxy would broaden the line spectrum, 
the observed electron spectrum could still have distinctive features. 
Thus, the observations of high-energy electrons 
bring us unique information about sources and propagation of cosmic rays, 
and enable us to search for dark matter.

Although the cosmic-ray electrons have been observed with many kinds of detectors 
since 1960 \citep{earl61},  
most observations are limited below several 100~GeV 
\citep{daniel65, golden84, tang84, grimani02, boezio00, duvernois01, 
torii01, aguilar02}. 
Among these observations, 
the first-time cosmic-ray electron observation with nuclear emulsions 
was achieved by \citet{daniel65}. 
They indicated that nuclear emulsions are ideal for the detection 
of electrons among many background protons 
because of the excellent imaging capability 
with a high position resolution of 1~$\mu$m.

The reason for the difficulty of the electron observations is 
that the electron flux itself is very low and decreases with energy 
much more rapidly than that of protons 
because of the electro-magnetic energy loss. 
The electron energy spectra are represented by a power-law function 
with an index of $-3.0$ to $-3.3$, 
which is steeper spectra than the proton spectra with a power-law index of $-2.7$ 
\citep[and references therein]{haino04}. 
The flux of cosmic-ray electrons is ${\sim}1$~\% of the protons at 10~GeV, 
and decreases very rapidly with increasing energy 
to be ${\sim}0.1$~\% of the protons at 1~TeV \citep[e.g.]{yoshida08}. 
Therefore, 
there are few observations of the electrons in the TeV region, 
since we need a long duration exposure with a detector 
that has a large geometrical factor, enough thickness, 
and powerful background rejection powers.

\citet{chang08} performed ATIC-2 balloon experiment in Antarctica and 
reported the energy spectrum in the energy region from 20~GeV up to 3~TeV, 
whose instrument 
contains a deep, fully active, BGO calorimeter of 18 radiation lengths (r.l.). 
They indicated an excess of cosmic-ray electrons at energies of 
$300 - 800$~GeV, 
compared to a general electron spectrum calculated with 
the GALPROP~\citep{galprop10}. 
They discussed that the excess may indicate a nearby source of energetic electrons 
such as the annihilated electrons from dark matter particles. 
On the other hand, 
from the independent data analysis of ATIC-2 $+$ ATIC-4, 
\citet{panov11} reported the electron spectrum from 30~GeV to 1~TeV, 
and indicated 
that the electron spectrum in the region of the excess includes 
a fine structure with a number of narrow peaks.

\citet{torii08} also observed cosmic-ray electrons from 10~GeV to 800~GeV 
by a long duration balloon flight using Polar Patrol Balloon (PPB) in Antarctica. 
The PPB-BETS is an imaging calorimeter composed of scintillating-fiber belts 
and plastic scintillators inserted between lead plates with 9 r.l. 
They discussed that 
the energy spectrum with PPB-BETS 
may indicate a sign of a structure in the several 100~GeV region, 
which is similar with the ATIC-2 observations, 
although a single power-law spectrum is acceptable within statistical errors. 

\citet{ackermann10b} presented the results of cosmic-ray electron observations 
from 7~GeV to 1~TeV 
using about $8{\times}10^{6}$ electron candidates detected 
in the first 12 months on-orbit by the Fermi Large Area Telescope (Fermi-LAT). 
Their electron spectrum can be described with a power law of 
$\propto E^{-3.08{\pm}0.05}$ with no prominent features, 
accommodating a slight spectral hardening at around 100~GeV and 
a slight softening above 500~GeV. 
Fermi-LAT also searched for 
anisotropies of electrons from 60 GeV to 480 GeV 
with angular scale extending from $\sim10^{\circ}$ to $90^{\circ}$, 
resulting in nul results \citep{ackermann10a}. 
They indicated that the upper limits for a dipole anisotropy range 
from ${\sim}0.5$~\% to ${\sim}10$~\%.
Although the Fermi-LAT has the large exposures of the electron observations, 
the detector thickness is insufficient to observe electrons in the TeV region. 
As the result, Fermi-LAT cannot separate electrons and protons one by one, 
but separated electrons from protons statistically 
based on Monte-Carlo simulations and machine learning algorithms.

The H.E.S.S. ground-based imaging atmospheric Cherenkov telescopes 
measured the electron spectrum 
in the energy range of 340~GeV to 5~TeV \citep{aharonian08, aharonian09}. 
The H.E.S.S. data show no indication of a structure 
in the electron spectrum, 
but rather a power-law spectrum with a spectral index of $-3.0$ 
which steepens to be around $-4.0$ above ${\sim}1$~TeV. 
While H.E.S.S. team reported electron observations up to several TeV, 
the electron spectrum is provided by indirect observations. 
Thus, H.E.S.S. intrinsically has systematic errors 
on the reconstructed electron spectra 
arising from uncertainties in the simulation of hadronic interactions, 
the atmospheric model, and the absolute energy scale.

\citet{adriani09} reported a statistically significant increase 
in the positron fraction at energies above ${\sim}10$~GeV 
with the PAMELA satellite-borne experiment, 
which is completely inconsistent with standard models 
describing the secondary production of cosmic rays. 
The PAMELA positron data indicate the existence of primary positron sources 
such as the annihilation of dark matter particles in vicinity of our Galaxy, 
nearby pulsars, and nearby micro-quasars. 
\citet{adriani11a} also presented 
the negatively charged cosmic-ray electron spectrum between 1 and 625~GeV 
performed by PAMELA, 
which is the first time that cosmic-ray negative electrons 
have been identified separately from positrons above 50~GeV. 
The negative electron spectrum can be described with a single power law energy 
dependence with a spectral index of $-3.18{\pm}0.05$ above 30~GeV 
and no significant spectral features.

We have observed high-energy cosmic-ray electrons from 30~GeV to 3~TeV with 
emulsion chambers at balloon altitudes, from 1968 to 2001, 
accumulating a total exposure of 8.19~m$^2$.sr.day. 
In the observations, we have carried out particle identification 
one event by one event with a proton rejection power larger 
than $1{\times}10^{5}$ in TeV region, 
because of an excellent imaging detector with a position resolution of 1~$\mu$m, 
which is one of the outstanding capabilities of the emulsion chambers. 
The performance of the emulsion chambers was examined with accelerator beam tests 
at CERN-SPS and Monte-Carlo simulations. 
We also estimated the atmospheric electron spectra 
in a reliable way \citep{komori12}, 
and carried out on-board calibrations by using the flight data.

While we previously reported the results from 1968 to 1976 
experiments \citep{nishimura80} 
and some additional publications (see references of \citet{kobayashi04}), 
in this paper we present the final cosmic-ray electron spectrum 
in the energy range from 30~GeV to 3~TeV 
observed with the balloon-borne emulsion chambers up to 2001, 
combined with our previous results.

\section{Detector}
\label{sec:detector}

\begin{figure}
\epsscale{0.80}
\plotone{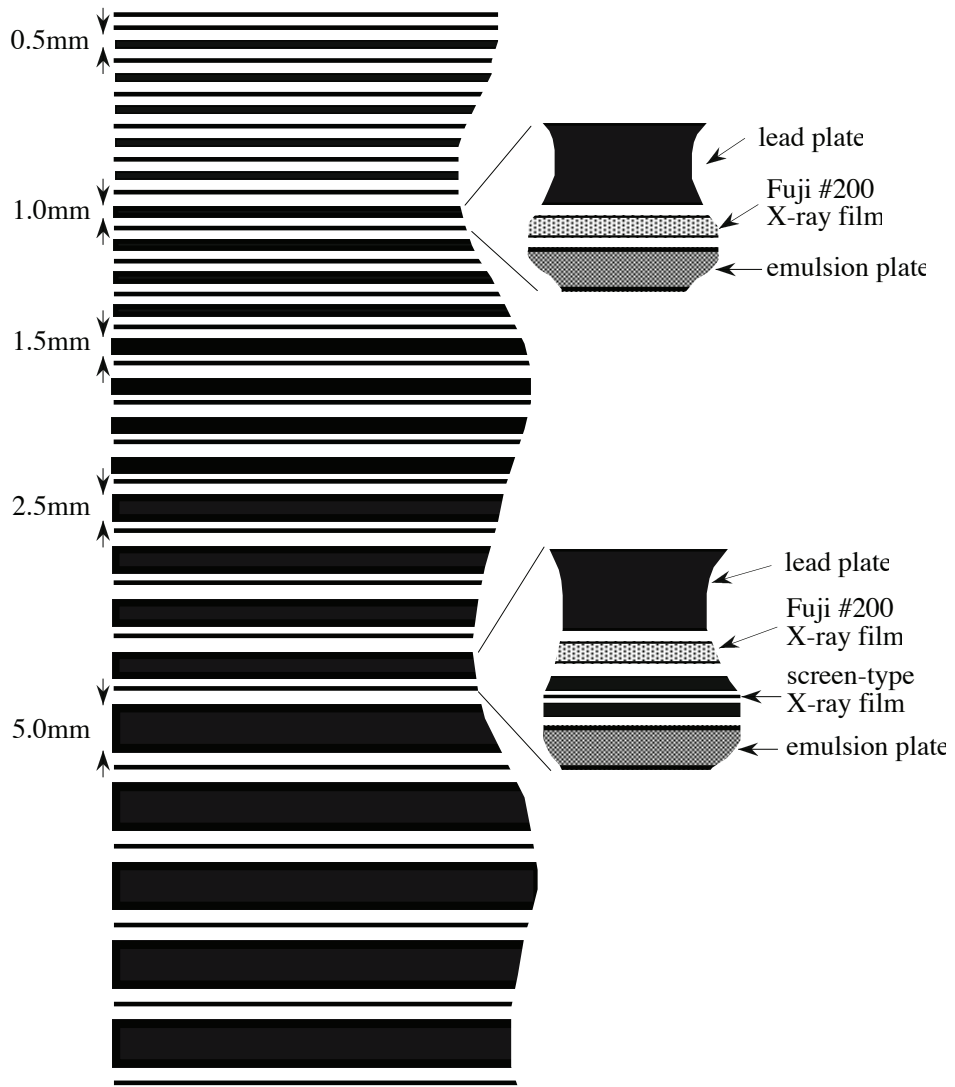}
\caption{
Typical configuration of the emulsion chamber in cross-sectional drawing 
from side view. 
\label{fig:ecc_config}}
\end{figure}

Emulsion chambers consists of nuclear emulsion plates, X-ray films, and 
lead plates (or tungsten plates in a few chambers). 
A nuclear emulsion plate is a methacrylate base 
$500-800$~$\mu$m thick, 
double coating of nuclear emulsion with $50-100$~$\mu$m thickness. 
We used Fuji ET-7B and ET-7D for nuclear emulsion. 
Nuclear emulsion plates are placed under lead plates. 
One or two X-ray films are inserted between a lead plate  
and a nuclear emulsion plate 
to allow rapid, naked-eye scanning for high-energy cascade showers, 
which produce dark spots in the films. 
Figure \ref{fig:ecc_config} shows a typical emulsion chamber configuration. 
The typical size and thickness of the detector are 
40~cm $\times$ 50~cm and 8~cm ($\sim9$~r.l.), respectively. 
Detailed configurations are described in \citet{nishimura80}.

The thickness of one lead plate at the upper layers is 0.5~mm ($\sim$0.09~r.l.) 
to identify incident parent particles, 
determine the incident angles, and 
investigate the initial shower developments. 
At the bottom layers, 
the thickness of one lead plate is 5~mm ($\sim$0.9~r.l.) and 
X-ray films are inserted to detect cascade showers, 
as shown in Fig.~\ref{fig:ecc_config}.

\begin{figure}
\epsscale{0.90}
\plotone{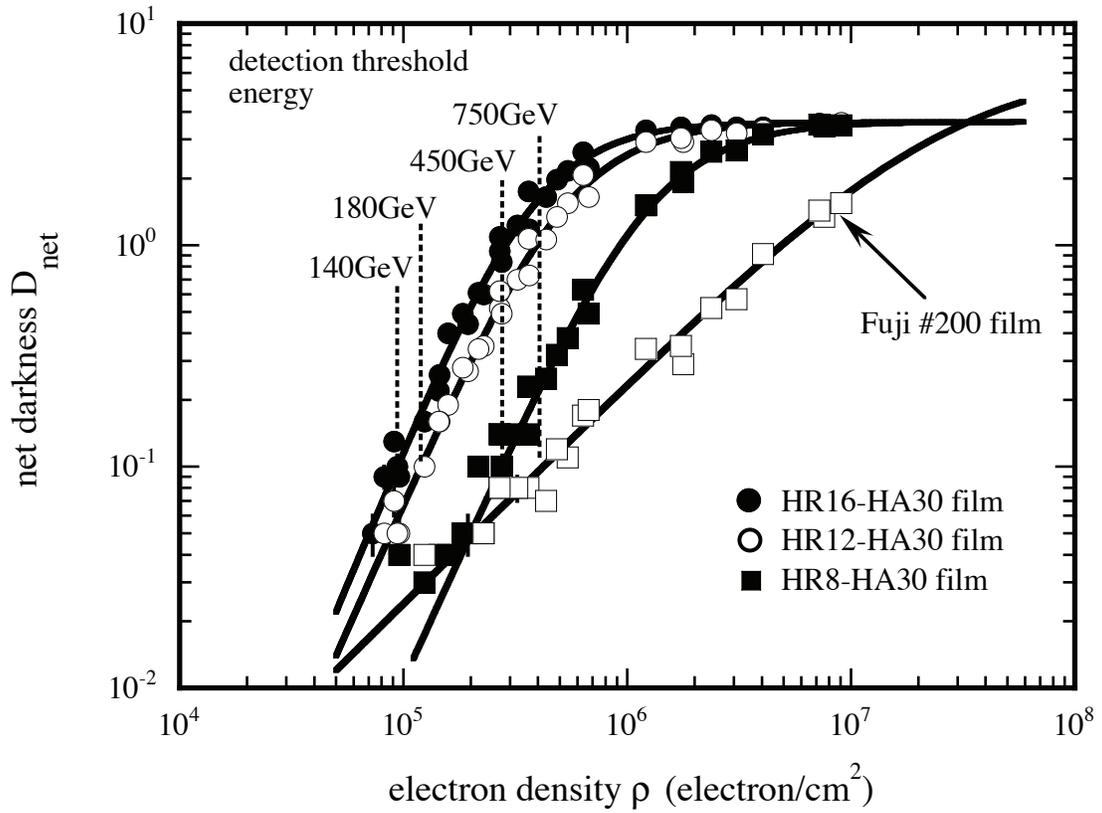}
\caption{
Characteristic curves for different types of X-ray films. 
\label{fig:x-ray_film}}
\end{figure}

Since high-energy electro-magnetic showers above a few 100~GeV 
leave dark spots on X-ray films, 
these showers can be detected with the naked eye by scanning the X-ray films. 
The corresponding tracks in the adjacent emulsion plate are located 
by using microscopes. 
The detection threshold of the X-ray film is 
$500$~GeV for Sakura type-N X-ray film used before 1984, 
$750$~GeV for Fuji {\#}200 X-ray film, 
and $250$~GeV, $200$~GeV, $150$~GeV 
for screen type X-ray films of Fuji G8-RXO, G12-RXO, and GS-RXO 
used from 1984 to 1988 \citep{kobayashi91}. 
After 1988 we used screen type X-ray films of 
HR8-HA30, HR12-HA30, and HR16-HA30. 
%
The sensitivity experiment of screen type X-ray films of HR series 
were carried out at Research Center for Electron Photon Science of 
Tohoku University in 2001 
using test chambers with multilayers of emulsion plates and X-ray films. 
The test chambers were exposed to the 200~MeV electron beams. 
Figure~\ref{fig:x-ray_film} shows the result of the characteristic curves 
of different type of X-ray films. 
Since the detection threshold of the net darkness on the X-ray films 
with naked eye is 0.1 \citep{kobayashi91}, 
the electron densities on the emulsion plates 
at the X-ray film detection threshold correspond to 
$0.9{\times}10^{5}$~cm$^{-2}$ for HR16-HA30, 
$1.2{\times}10^{5}$~cm$^{-2}$ for HR12-HA30, 
$2.6{\times}10^{5}$~cm$^{-2}$ for HR8-HA30, 
and 
$4.0{\times}10^{5}$~cm$^{-2}$ for Fuji {\#}200 X-ray films. 
These electron densities are compatible with 
the shower track densities with emulsion chambers 
at the shower maximum of electrons with energies of 
140~GeV, 180~GeV, 450~GeV, and 750~GeV, respectively 
(see Fig.~\ref{fig:ecc_trcv}), 
which are the detection threshold energies of the X-ray films. 

\begin{center}
\vspace*{0.5cm}
\framebox[3cm]{Figure \ref{fig:ecc_config}}
\vspace*{0.5cm}
\end{center}

\begin{center}
\vspace*{0.5cm}
\framebox[3cm]{Figure \ref{fig:x-ray_film}}
\vspace*{0.5cm}
\end{center}

Because of the simple configuration of the detector, 
the geometrical factor ($S{\Omega}$) can be estimated very accurately, 
a difficult task for some electronic detectors. 
For the electron observations, 
the effective geometrical factor is given by 
\begin{equation}
S{\Omega}_{e} = 2{\pi}S{\eta}\int_0^{\theta_0} {\cos}{\theta}{\sin}{\theta}
d{\theta} = {\pi}S{\eta}{\sin}^2{\theta_0}, 
\end{equation}
where ${\theta_0}$ is the upper limit of incident angles and 
$\eta$ is so-called "edge effect". 
Since the detector has its edge, 
some incident electrons near the edge on the top of the detector 
do not pass through the bottom of the detector. 
The edge effect is the efficiency of events that pass through the top and bottom 
emulsion plates. 
In the typical case of $\theta_0 = 60^{\circ}$ and $S=0.40{\times}0.50$~m$^2$, 
$S{\Omega}_{e}$ is 0.39~m$^2$~sr with $\eta$ of 0.82 
for the chamber thickness of 8.0~cm. 
In appendix \ref{sec:ecc_list}, 
we summarize the area $S$, edge effect $\eta$, 
upper limit of incident angles ${\theta}_{0}$, 
and $S{\Omega}_{e}T$ of the emulsion chambers. 
As shown in appendix \ref{sec:ecc_list}, 
$S$, $\eta$, ${\theta}_{0}$, and $S{\Omega}_{e}T$ 
change depending on the electron energies. 

We measure the shower particles within a circle of $100$~$\mu$m radius 
from shower axis. 
This means that we select the shower particles with higher energies, 
which suffered less multiple scattering in the chamber. 
Hence, the number of the shower particles selected decreases faster than 
that of all shower particles. 
The shower maximum in emulsion chambers for the shower particles 
within a circle of $100$~$\mu$m radius 
appears in $\sim6$~r.l. for 1~TeV electrons, 
while the maximum of the total number of shower particles 
appears in $\sim12$~r.l. for 1~TeV electrons. 
As a result, the energy of higher energy incident electrons 
can be determined with a thinner detector. 
Thus the emulsion chamber has the advantages of a wide field of view, 
small thickness, and lightweight detector, compared to the other instruments.

\begin{figure}
\epsscale{0.80}
\plotone{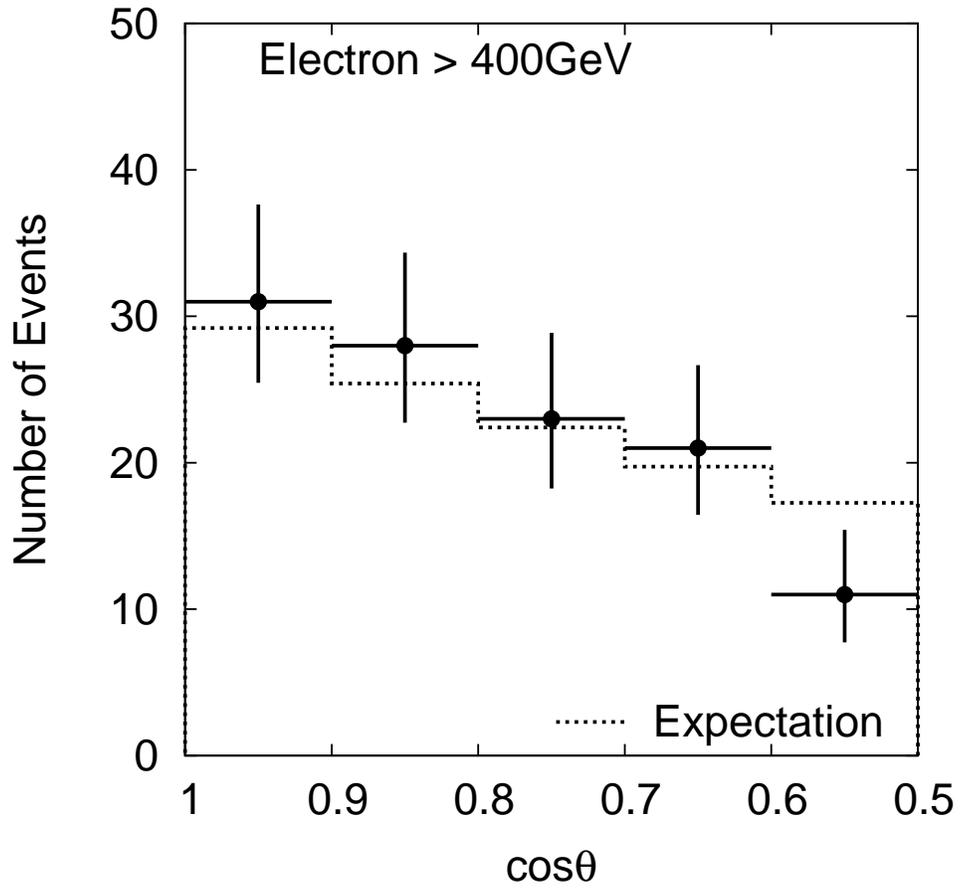}
\caption{
Zenith angle distributions for electrons above 400~GeV 
observed with the balloon-borne emulsion chambers, 
which is compared to the expected distribution for primary electrons. 
\label{fig:zenith}}
\end{figure}

In order to verify the zenith angle dependence of the detection efficiency 
for the incident electrons, 
we present the zenith angle distribution of electrons 
observed with the balloon-borne emulsion chambers in Fig.~\ref{fig:zenith}, 
which is compared to the expected distribution for primary cosmic-ray electrons. 
As shown in Fig.~\ref{fig:zenith}, 
the zenith angle distribution of electrons 
is consistent with the expectation.

\begin{center}
\vspace*{0.5cm}
\framebox[3cm]{Figure \ref{fig:zenith}}
\vspace*{0.5cm}
\end{center}

\section{Balloon observations}
\label{sec:observations}

\begin{figure}
\epsscale{0.90}
\plotone{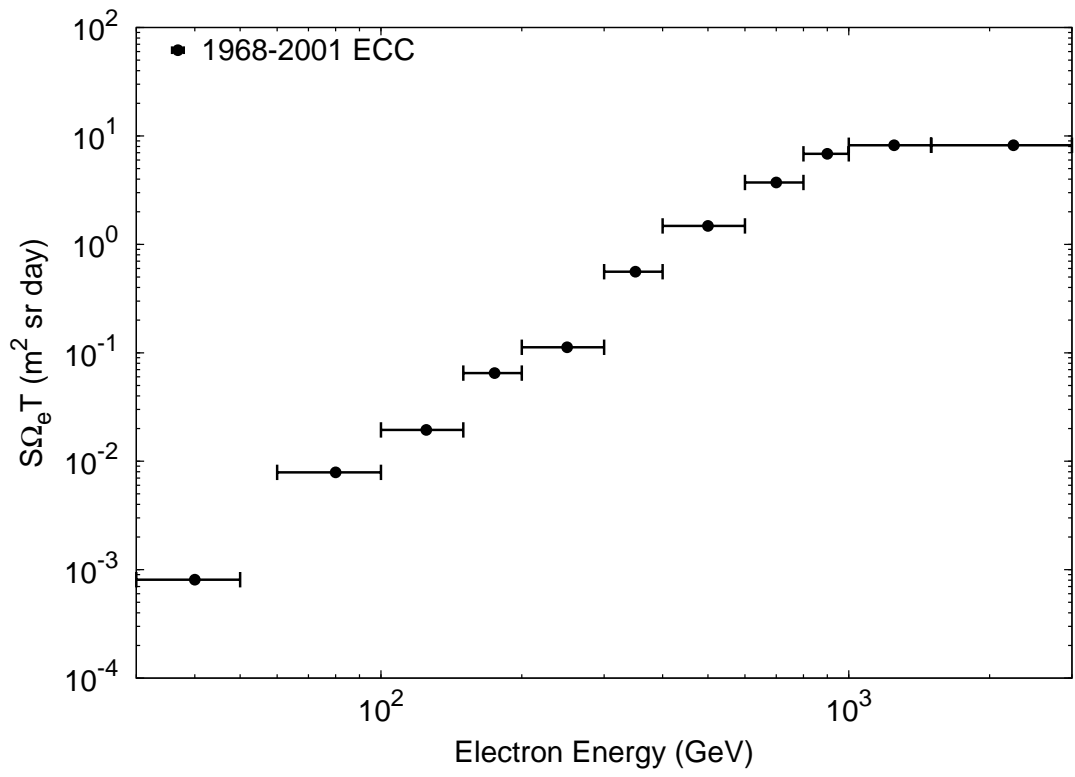}
\caption{
Total exposure $S{\Omega}_{e}T$ for primary electrons with electron energy. 
\label{fig:exposure}}
\end{figure}

We have observed cosmic-ray electrons with balloon-borne emulsion chambers 
in 14 flights between 1968 and 2001. 
In order to reject background cosmic rays, 
the emulsion chambers are placed upside down 
in the balloon gondolas during ascent and descent of the balloons, 
and are flipped to a normal position during level flight. 
The pressure altitude records for each flight 
correspond to residual atmospheric overburdens 
in the range from 4.0~g~cm$^{-2}$ to 9.4~g~cm$^{-2}$. 
In Table \ref{tab:baln_list}, 
we summarize the series of experiments since 1968, 
in which the results for 1968 -- 1976 observations were reported 
in \citet{nishimura80}. 
The $S{\Omega}_{e}T$ in Table \ref{tab:baln_list} 
present the effective exposure factors for primary electron observations 
in the energy range above 1~TeV within the zenith angle of $60^{\circ}$. 
Figure~\ref{fig:exposure} shows 
the total cumulative effective exposure $S{\Omega}_{e}T$ 
for primary electrons, which is 8.19~m$^2$-sr-day in the TeV region.  
In addition to the electron observations, 
we have simultaneously observed atmospheric gamma rays, 
whose results are described in \citet{yoshida06}.

\begin{center}
\vspace*{0.5cm}
\framebox[3cm]{Table \ref{tab:baln_list}}
\vspace*{0.5cm}
\end{center}

\begin{center}
\vspace*{0.5cm}
\framebox[3cm]{Figure \ref{fig:exposure}}
\vspace*{0.5cm}
\end{center}

\section{Data Analysis}
\label{sec:analysis}

In the balloon observations, we identify electron events 
among incoming cosmic-ray events and determine the energies. 
In the following data analysis, 
we selected events with 
incident zenith angle less than 60 degree 
and which passed all the way from the top to bottom layer of the chambers.

\subsection{Event identification}

High-energy electro-magnetic showers above a few 100~GeV 
are detected with naked-eye scanning of dark spots 
left on the X-ray films. 
The corresponding tracks in the adjacent emulsion plate 
are located with microscopes, and 
traced back through the stack to the shower starting points. 
%
As described in section \ref{sec:detector} and appendix \ref{sec:ecc_list}, 
since we have improved the X-ray films to detect lower energy electrons, 
we have used different types of X-ray films 
that have the different threshold energies. 
Hence, 
the total cumulative effective exposure $S{\Omega}_{e}T$ 
depends on electron energies, as shown in Fig.~\ref{fig:exposure}. 
The detection efficiencies are 100~\% above the threshold energies
and fall off rather rapidly below the thresholds \citep{nishimura80}. 
We also confirmed the threshold energy for each balloon flight 
by using the deviation from a single power-law spectrum of 
the observed atmospheric gamma rays. 
In this analysis, 
we used the electron events above the threshold energies 
to derive the electron energy spectrum. 
%
In order to detect the electro-magnetic showers below a few 100~GeV, 
the emulsion plates were directly scanned with microscopes 
for a part of the 1968, 1969, 1970, 1973, and 1996 emulsion chambers. 
We successfully detected electron events down to 30~GeV 
with the microscope scanning. 
The detection efficiency is larger than 95\% \citep{nishimura80}. 
The microscope scanning is carried out in the smaller rectangular area 
of the upper emulsion plates. 
Since the shower particles on the bottom emulsion plates are measured 
in the full area, 
the edge effect $\eta$ of the microscope scanning has the larger value than 
that of the usual shower measurements. 
In the case of the scanning area 
within $d {\tan} {\theta}_{0}$ from the detector edge, 
where $d$ is the thickness of the detector 
and ${\theta}_{0}$ is the upper limit of incident angles, 
$\eta$ is $1.00$. 
For the larger scanning area, 
not within $d {\tan} {\theta}_{0}$ from the detector edge, 
$\eta$ is smaller than $1.00$, as shown in Table~\ref{tab:ecc_list}. 
In emulsion chambers, it is possible to measure the location of 
shower tracks in each emulsion plate with a precision of $1$~$\mu$m. 
The incoming particles 
such as electrons, gamma-rays, protons, and heavier nuclei 
are identified by examining the details of shower development, 
especially around the shower starting points.

Since electron events start from a single charged track 
which produces an electron-positron pair within 1~r.l. of the top of the emulsion chamber 
with about 90~\% probability, 
they are identified by the existence of a single and a pair track 
with the spreading angle less than $1{\times}10^{-3}$~rad at the interaction point, 
as described in Appendix \ref{sec:rejection}. 
Electron events also give the electro-magnetic shower 
without core structures. 
Gamma-ray events, which are also a pure electro-magnetic shower, 
start from a pair with no visible primary track 
above the shower starting point. 
Although the incident track of a proton-induced shower 
shows a single charged track like an electron, 
proton-induced showers also have many secondaries at the shower starting point 
and often have multi core structures in the deep layers. 
Even in the case of proton-induced showers with few secondary tracks, 
it is possible to discriminate the proton-induced showers from electron-induced showers 
by the differences of the spreading angle between tracks 
at the shower starting points. 
As described in Appendix \ref{sec:rejection}, 
the proton rejection power is estimated to be larger than $1{\times}10^5$, 
that is derived to be independent of M.C. simulation codes and 
hadron interaction models.
Hadron showers of heavier nuclei such as helium 
are easily distinguished because the grain density of the incident track
is larger than a minimum ionizing particle.

\begin{figure}
\epsscale{0.49}
\plotone{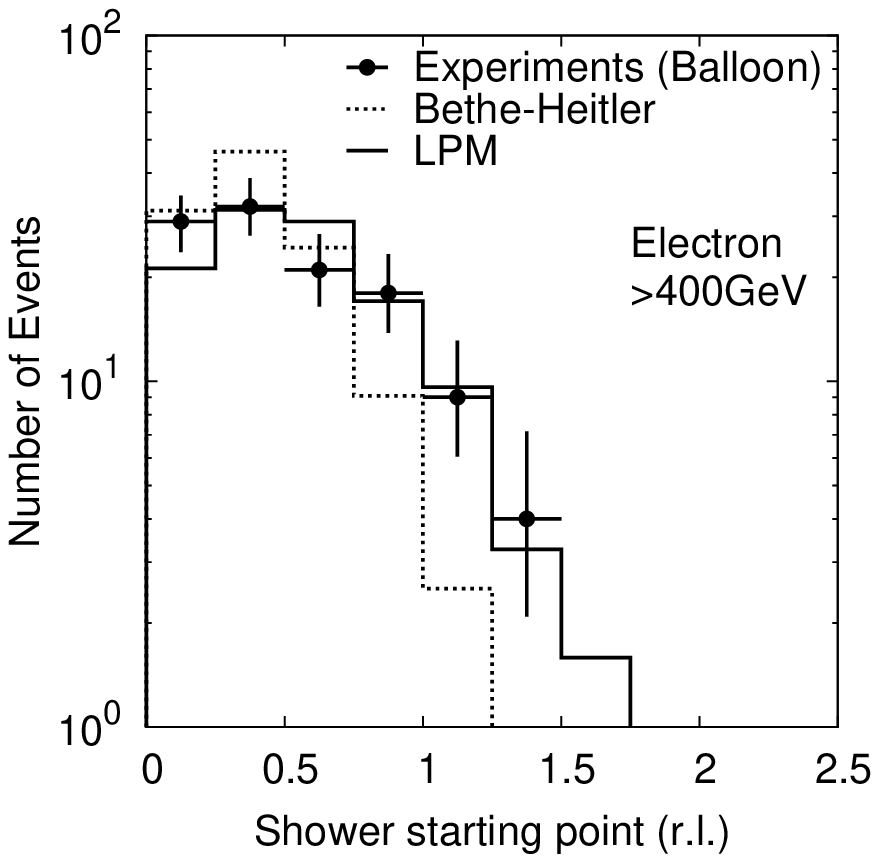}
\plotone{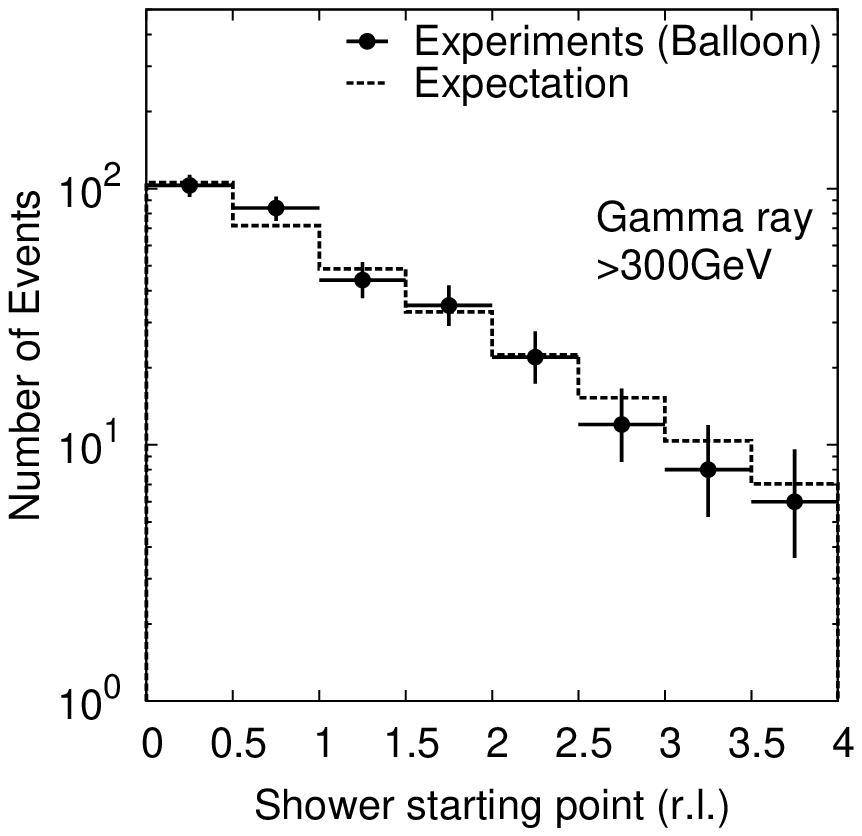}
\plotone{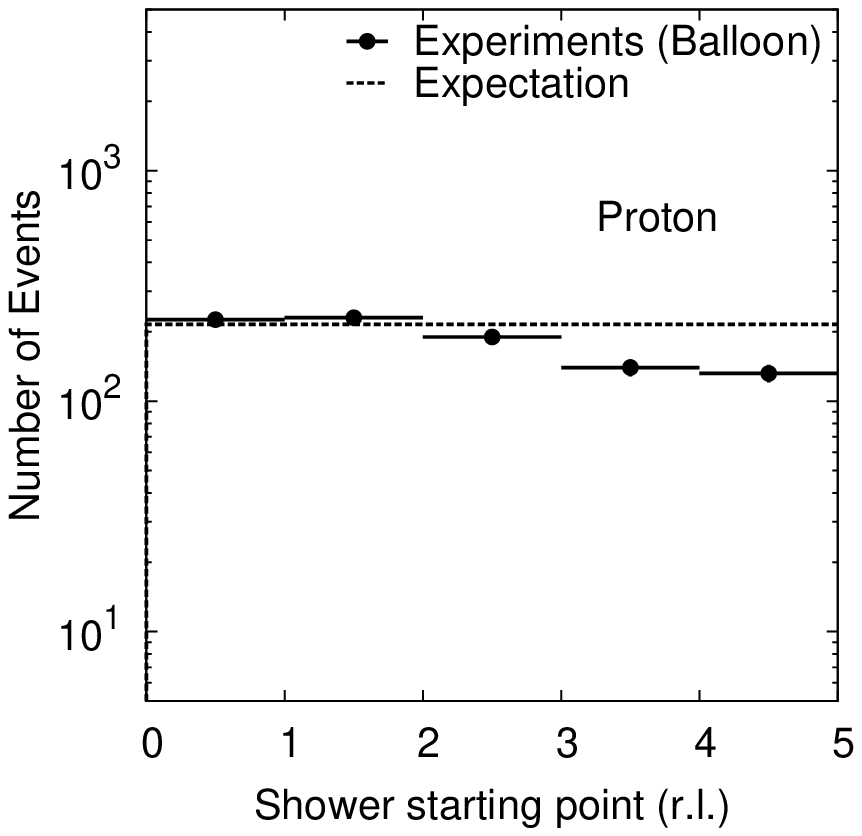}
\caption{
Shower starting point distributions for the observed electrons, 
gamma rays, and protons, compared to the expected distributions. 
\label{fig:stp}}
\end{figure}

In emulsion chambers, 
we can measure the depth of the first electron-positron pair 
of the electron-induced shower, the so called shower starting point. 
The validity of event identification can be checked 
by comparison of the measured shower starting points 
with the expected values. 
Figure~\ref{fig:stp} presents the shower starting point distributions 
of the balloon observations 
for electrons above 400~GeV, gamma rays above 300~GeV, and protons, 
compared to the expected distributions. 
As for the electrons, 
the shower starting point is compared to the Bethe-Heitler expectation 
and the LPM expectation based on Migdal's formula \citep[e.g.]{baier05}. 
As shown in Fig.~\ref{fig:stp}, 
the shower starting point distributions within 3.0~r.l. 
show good agreement with the expectations, 
whose results suggest the reliability of the particle identification, 
and the deviation of the proton distribution larger than 3.0~r.l. 
from the expectation shows the decrease of the proton detection efficiency. 
In particular, 
the consistent result of the electrons with the LPM expectation 
strongly suggests the accurate identification of electrons.

\begin{center}
\vspace*{0.5cm}
\framebox[3cm]{Figure \ref{fig:stp}}
\vspace*{0.5cm}
\end{center}

\subsection{Energy Determination}

Electron energies were determined by counting the number of shower 
tracks in each emulsion plate 
within a circle of radius 100$\mu$m centered on the shower axis. 
We derived the integral track length from these 
counted tracks in each layer. 
The integral track length is theoretically expected to be proportional 
to the shower energy, as discussed in detail in \citet{nishimura80}. 
Our chamber structures are slightly different in each flight 
because of slight differences of lead thicknesses and insertion of different
types of X-ray films and phosphoric screen films. 
Since the differences of the chamber structure affect the integral track lengths, 
we calculated the shower developments for each chamber one by one 
using a Monte-Carlo simulation code called Epics \citep{kasahara12}. 
Epics has been used for cosmic-ray experiments \citep[e.g.]{torii01, amenomori09}, 
and also used for very forward single photon energy spectra 
from 0.1~TeV to 3.6~TeV in the Large Hadron Collider forward (LHCf) experiments \citep{adriani11b}. 
The incident electron energies are determined by these  
track lengths compared with 
the values estimated from the Monte-Carlo simulations for each chamber.

\begin{figure}
\epsscale{1.00}
\plotone{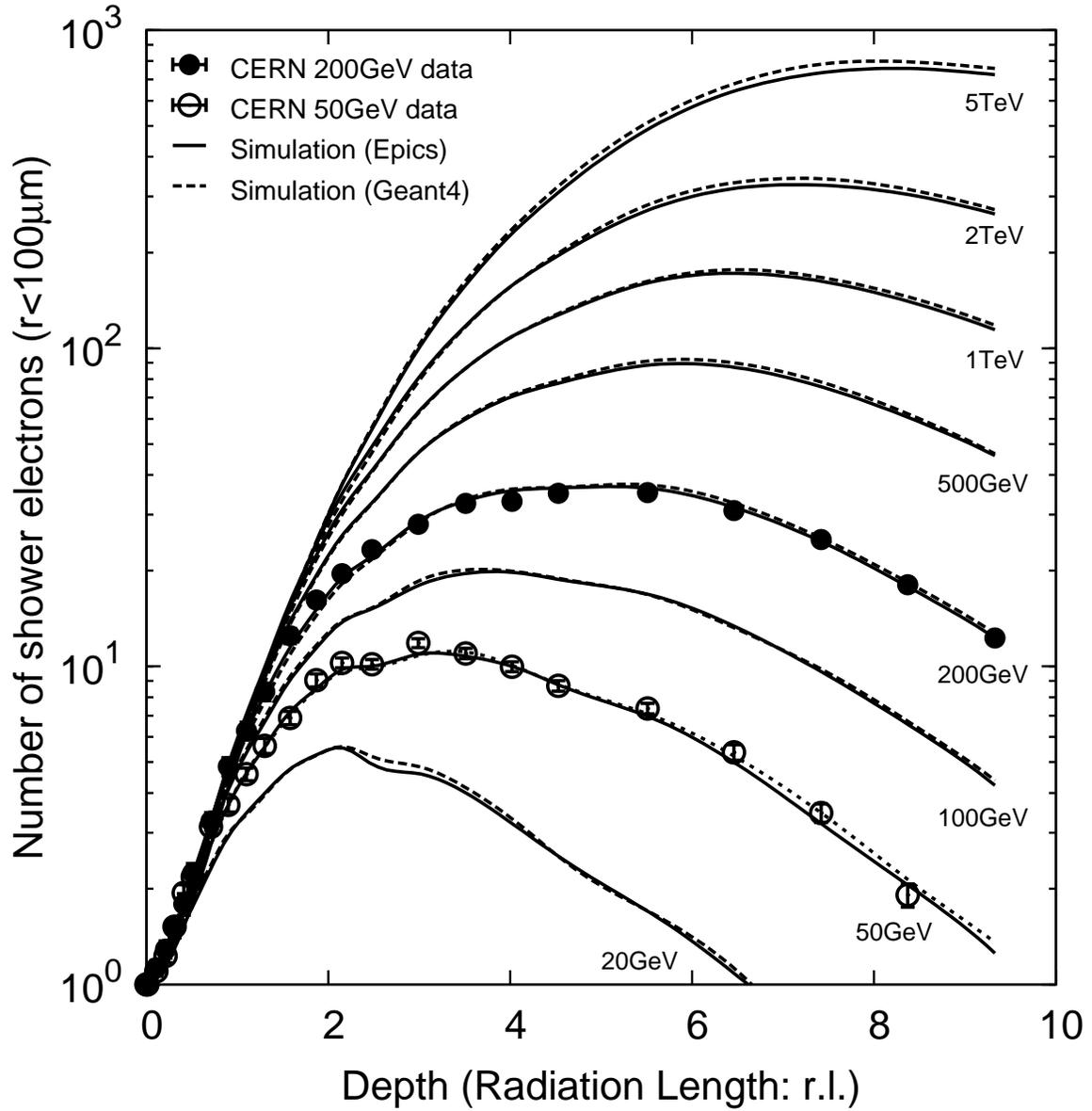}
\caption{
Longitudinal developments of the averaged number of shower tracks 
within a radius of 100~$\mu$m from the simulations, 
compared to the experimental data. 
\label{fig:ecc_trcv}}
\end{figure}

\begin{figure}
\epsscale{0.65}
\plotone{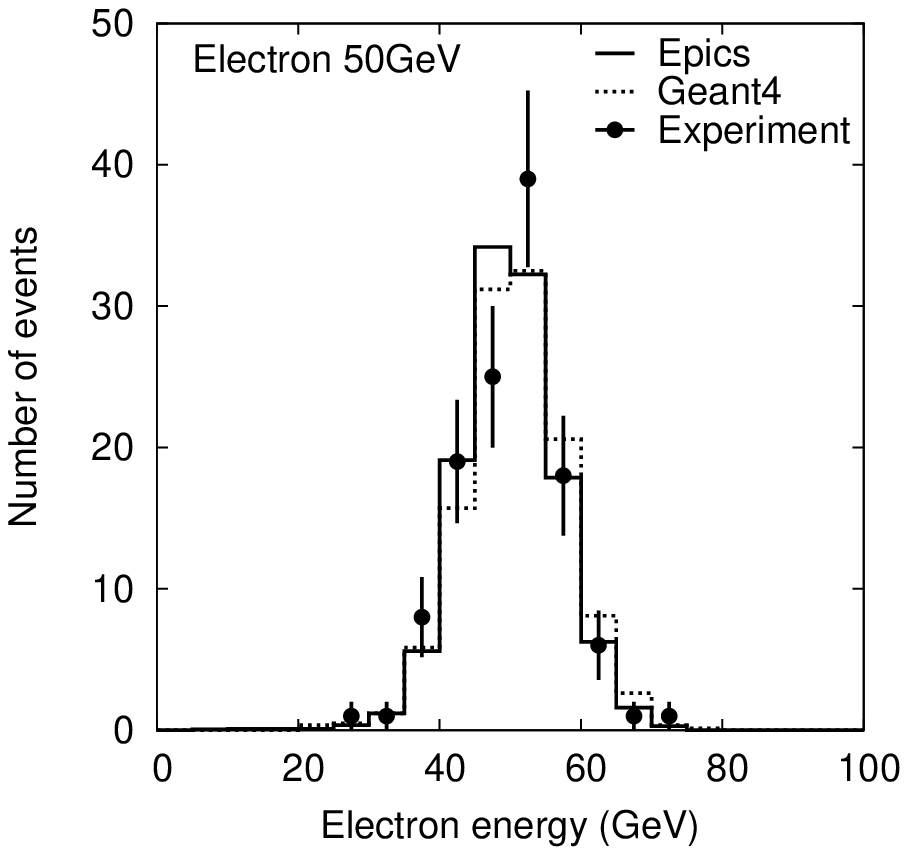}
\plotone{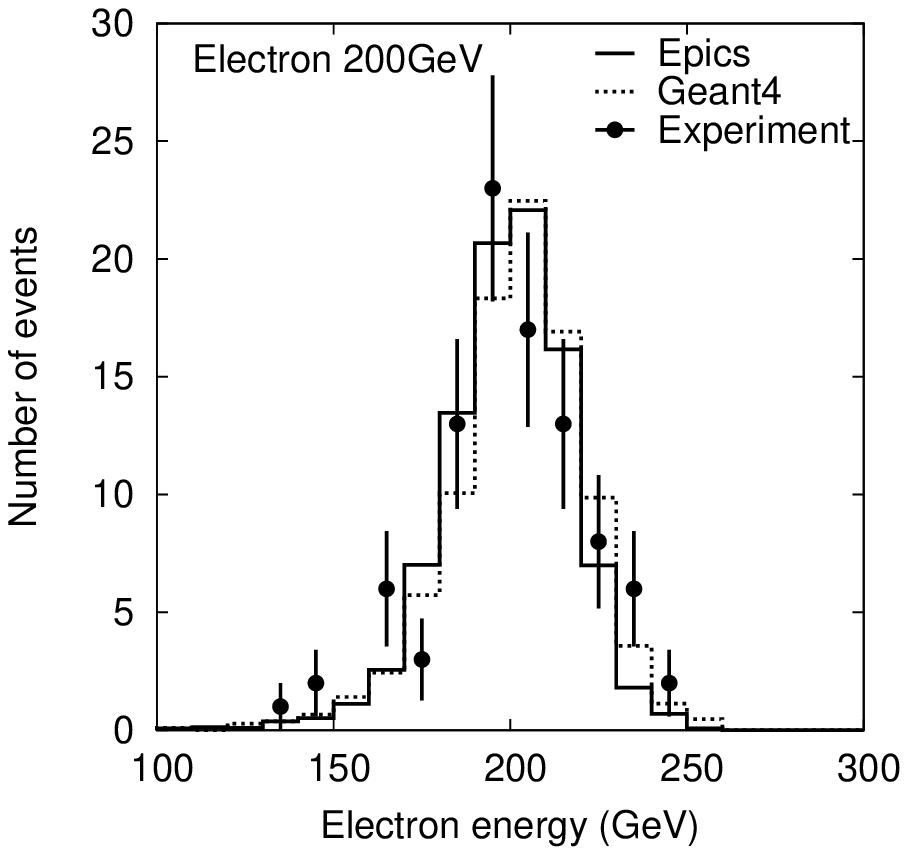}
\caption{
Energy distributions of the experimental data for 50~GeV 
and 200~GeV electron beams at CERN-SPS, compared with the simulations. 
\label{fig:TL_hist}}
\end{figure}

\begin{figure}
\epsscale{0.80}
\plotone{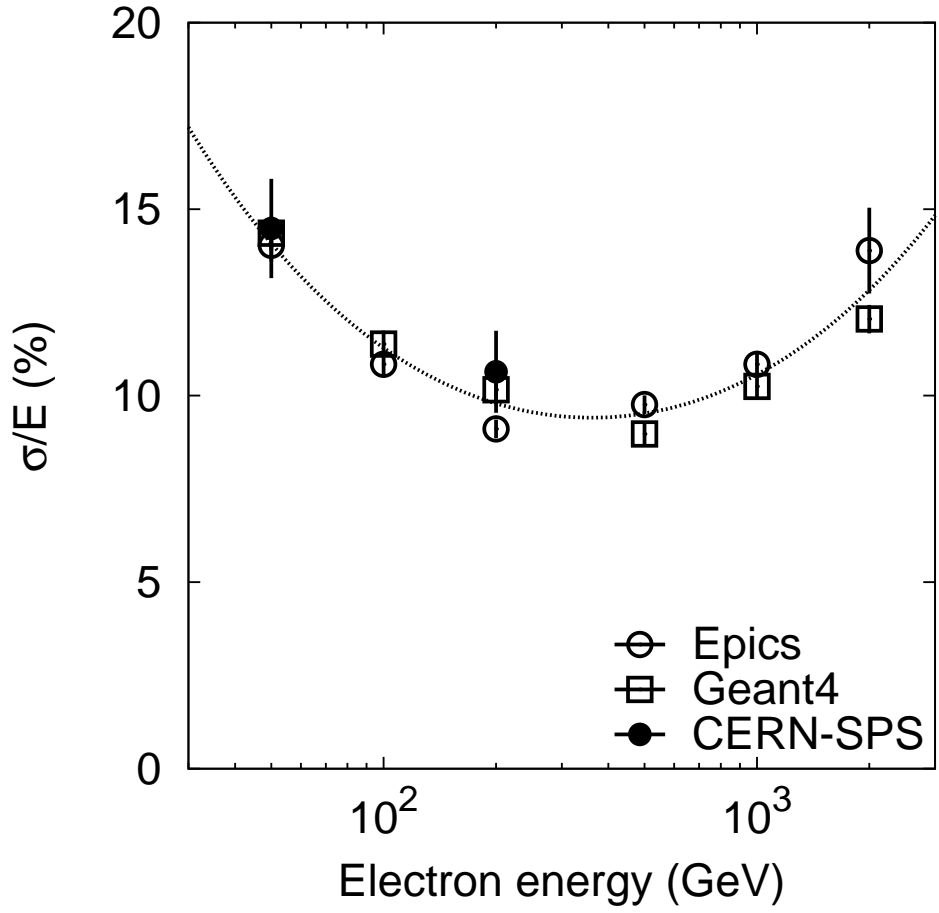}
\caption{
Energy dependence of energy resolutions with the emulsion chambers 
from the simulations, compared to the experimental data for electrons 
of 50~GeV and 200~GeV. 
\label{fig:ecc_eneres}}
\end{figure}

For the calibration of the detector, 
we carried out beam tests of electrons in 2004 at CERN-SPS. 
The detector configuration is same as the balloon-borne emulsion chambers, 
except for the detector size of 10.0~cm ${\times}$ 12.5~cm.
Results calculated using the Epics code were confirmed by 
emulsion chambers exposed to 50~GeV and 200~GeV electron beams at CERN-SPS. 
%
In order to evaluate the possible systematic errors at energies greater than 200~GeV, 
we also compared two independent Monte-Carlo simulation codes; 
Epics and Geant4 \citep{agostinelli03, amako06}. 
Figure~\ref{fig:ecc_trcv} shows 
the longitudinal development of the average number of shower tracks 
from the Monte-Carlo simulations, 
compared with the results of 50~GeV and 200~GeV electron beams. 
As shown in Fig.~\ref{fig:ecc_trcv}, 
the simulations well represent the experimental data. 
The differences between the integral track lengths of Epics and Geant4 are ${\sim}2$~\% 
in the energy range of 30~GeV -- 3~TeV, 
which is negligible small compared to statistical errors 
of our cosmic-ray electron spectrum as described in section \ref{sec:spectrum}. 
Figure~\ref{fig:TL_hist} shows the energy distributions 
for 50~GeV and 200~GeV electron beams.  
The determined energies with the simulations 
for 50~GeV and 200~GeV electrons are consistent with the experimental data. 
The energy resolutions are 14.5~\% at 50~GeV and 10.6~\% at 200~GeV, respectively. 
Figure~\ref{fig:ecc_eneres} presents 
the energy resolutions of the simulations, 
compared with the experimental data. 
The energy resolution for the emulsion chamber 
is well represented by the form of 
\begin{equation}
\frac{\sigma}{E} = [ 8.6\%^{2}(\frac{E}{100{\rm GeV}})^{-1} 
 + 6.9\%^{2} + 2.4\%^{2}(\frac{E}{100{\rm GeV}}) ]^{1/2}, 
\end{equation}
where $E$ is the electron energy and 
$\sigma$ is the standard deviation of energy determination. 
The first term in right-hand side root represents statistics-related fluctuations 
of the number of shower particles, 
while the last term represents fluctuations due to shower particles 
escaping from the finite thickness of the detector.

\begin{center}
\vspace*{0.5cm}
\framebox[3cm]{Figure \ref{fig:ecc_trcv}}
\vspace*{0.5cm}
\end{center}

\begin{center}
\vspace*{0.5cm}
\framebox[3cm]{Figure \ref{fig:TL_hist}}
\vspace*{0.5cm}
\end{center}

\begin{center}
\vspace*{0.5cm}
\framebox[3cm]{Figure \ref{fig:ecc_eneres}}
\vspace*{0.5cm}
\end{center}

\subsection{Electron energy spectrum}
\label{sec:spectrum}

In balloon flight experiments, 
it is necessary to correct the observed cosmic-ray electron spectrum 
because of the residual overlying atmosphere. 
We corrected energy loss of primary electrons due to bremsstrahlung radiation 
in the overlying atmosphere. 
The average bremsstrahlung energy loss to each electron 
is given by 
\begin{equation}
E_{0} = E e^{ \frac{A(s)}{s} {\cdot} \frac{t}{{\cos}{\theta}} }, 
\end{equation}
when the incident electron spectrum is a power-law function 
with an index of $-(s+1)$. 
Here, 
$E_{0}$ is the energy of primary electron at the top of atmosphere, 
$E$ is the measured energy in the detector,  
$t$ is the vertical thickness of the overlying atmosphere in radiation lengths, 
$\theta$ is a zenith angle of the incident electron to the detector, 
and $A(s)$
refers to the function used in electro-magnetic shower theory 
\footnote{
$A(s)$ is given by as follows:  
\[
A(s) = 1.36\frac{ d {\log}{\Gamma}(s+2) }{ ds } - \frac{1}{(s+1)(s+2)} - 0.075. 
\] 
}, 
described by \citet{nishimura67}. 
In the case of $s = 2.3$, $A(2.3)$ is $1.674$. 
For example, in the case of a power-law index of $-3.3$ ($s=2.3$), 
an atmospheric thickness of $6.0$~g~cm$^{-2}$, 
and a zenith angle of $45$ degree, 
the energies of incident electrons are reduced by $16$~\%, 
and hence the electron flux decreases by $32$~\%. 
This energy loss formula is different from the simple energy loss of 
$E_{0} = E e^{t/{\cos}{\theta}}$, 
since the energy losses of electrons have broad distributions and 
the incident electron spectrum is steeply sloped. 
In the case of $E_{0} = E e^{t/{\cos}{\theta}}$ with the same parameters, 
the energies are reduced by $21$~\% 
and the electron flux decreases by $41$~\%, 
which correspond to monochromatic electrons.

In addition to primary cosmic-ray electrons, 
atmospheric electrons are also produced by hadronic interactions 
of primary cosmic rays with nuclei in the atmosphere. 
Since almost all atmospheric electrons are produced 
via atmospheric gamma rays from neutral pion decay, 
the atmospheric electron spectrum 
is estimated by using the simultaneously observed atmospheric gamma-ray spectrum 
with the emulsion chambers \citep{yoshida06}. 
\citet{komori12} derived the atmospheric electron spectrum 
in the upper atmosphere less than $10$~g~cm$^{-2}$ 
from the observed gamma-ray spectrum using the electro-magnetic shower theory. 
Their derived atmospheric electron spectrum is substantively free 
from the uncertainties of the cosmic-ray nuclear spectra 
and hadronic interaction models. 
The contributions of the atmospheric electrons to primary electrons 
increase with electron energies and with thicknesses of 
the overlying atmosphere. 
Table~\ref{tab:elec_flux} shows the number of atmospheric electrons, 
which ranges from $0$~\% to $50$~\% of the observed electrons.

We observed electrons at each balloon altitude, 
and derived the cosmic-ray electron spectrum using the following formula: 
\begin{equation}
J_{\rm e}(E) = 
\frac{ N_{\rm e} - N_{\rm 2nd} }{ S{\Omega}_{e} T {\Delta} E C_{\rm eff} C_{\rm enh} }
( {\rm m}^{-2} {\rm s}^{-1} {\rm sr}^{-1} {\rm GeV}^{-1} ) .
\end{equation}
Here, 
$N_{\rm e}$ is the number of the observed electron events, 
$N_{\rm 2nd}$ is the number of atmospheric electrons, 
$C_{\rm eff}$ is electron detection efficiency, 
$C_{\rm enh}$ is enhancement factor due to the energy resolution.

The efficiency $C_{\rm eff}$ for detecting electro-magnetic showers in emulsion chambers 
have been studied to be essentially $100$~\% above the threshold energy 
for naked-eye scanning of X-ray films 
\citep[and references therein]{kobayashi91}. 
The detection efficiency for HR16-HA30 
was also tested by using emulsion chambers 
exposed to the 200~GeV electron beam at CERN-SPS. 
Simultaneously with primary electrons, 
we have also observed atmospheric gamma rays to check the performance 
of the emulsion chambers in each balloon experiment \citep{yoshida06}. 
We also confirmed the detection efficiency of each emulsion chamber 
from the atmospheric gamma-ray spectra. 
The uncertainty of the energy determination has the effect 
of enhancing the absolute flux of electrons, 
in particular, for a steep power-law spectrum. 
The enhancement factor $C_{\rm enh}$ due to the energy resolution 
has values from 1.01 to 1.09, depending on electron energies 
(see \citet{yoshida06} in detail).

\section{Results and discussions}

\begin{figure}
\epsscale{1.00}
\plotone{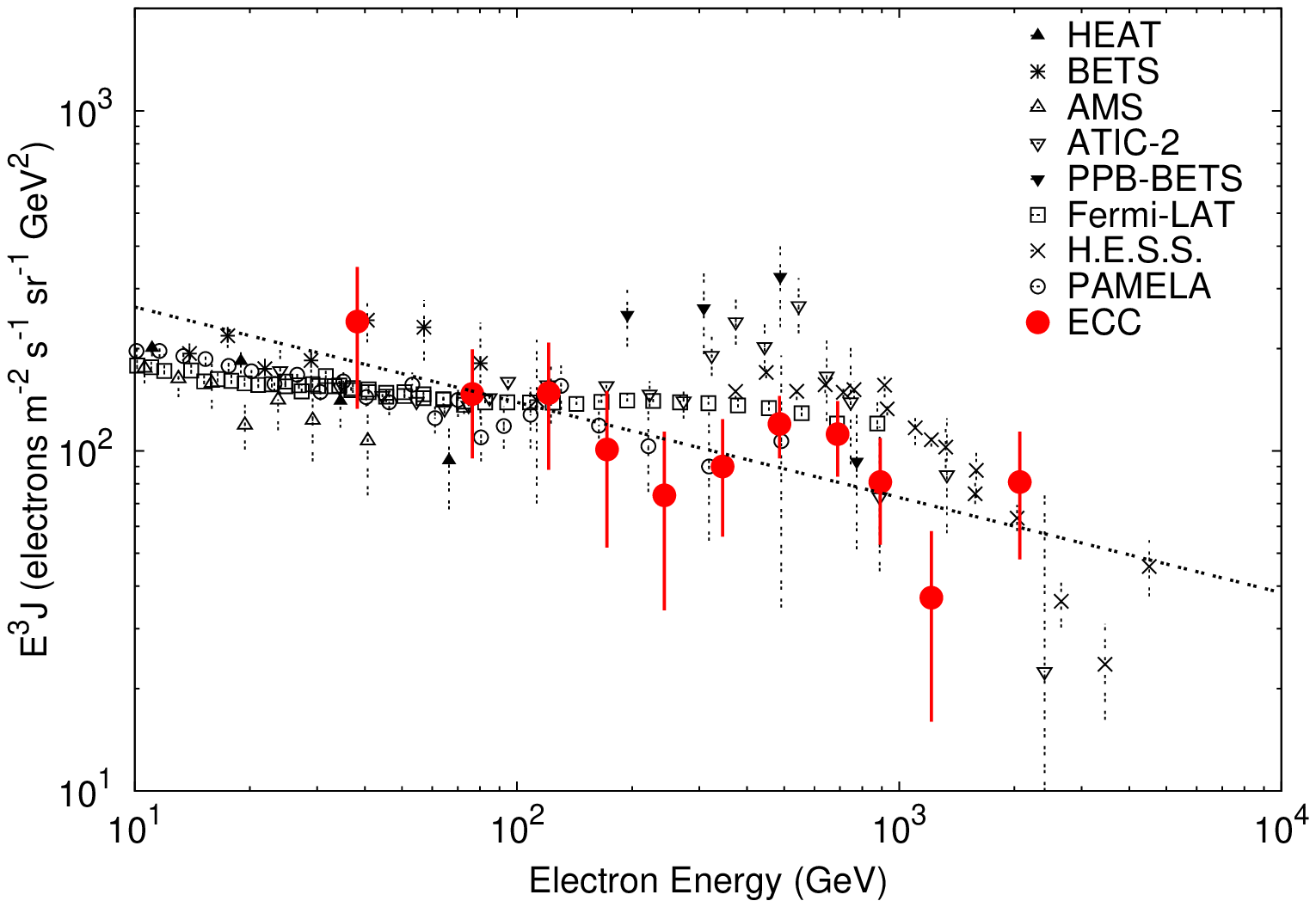}
\caption{
The primary cosmic-ray electron spectrum observed 
with the emulsion chambers (ECC), 
compared to the recent experiments 
\citep{duvernois01, torii01, aguilar02, chang08, ackermann10b, aharonian09, adriani11a}. 
The dotted line shows the best fitted power-law spectrum 
with an index of $-3.28$. 
\label{fig:espec}}
\end{figure}

The total number of the observed electrons is 166 events 
with the balloon-borne emulsion chambers exposed from 1968 to 2001 
in the energy range of 30~GeV to 3~TeV. 
After the corrections described above, 
we derived the primary cosmic-ray electron energy spectrum. 
Figure~\ref{fig:espec} shows the observed electron spectrum, 
which is well represented by a power-law function of 
\begin{equation}
J_{\rm e}(E) = (1.39{\pm}0.23){\times}10^{-4} (E/100{\rm GeV})^{-3.28{\pm}0.10} 
({\rm m}^{-2} {\rm s}^{-1} {\rm sr}^{-1} {\rm GeV}^{-1}). 
\end{equation}
The flux values and numbers of the electrons in each energy bin 
are listed in Table~\ref{tab:elec_flux}. 
Compared with our previous electron spectrum 
in the energy range of $30 - 1000$~GeV \citep{nishimura80},  
the total number of the observed electrons increased threefold the previous result, 
and the highest energy was extended up to 3~TeV.

\begin{center}
\vspace*{0.5cm}
\framebox[3cm]{Figure \ref{fig:espec}}
\vspace*{0.5cm}
\end{center}

\begin{center}
\vspace*{0.5cm}
\framebox[3cm]{Table \ref{tab:elec_flux}}
\vspace*{0.5cm}
\end{center}

The cosmic-ray electrons observed with balloon-borne emulsion chambers (ECC) 
extend up to 3~TeV with no cut off 
in the form of a power-law spectrum with an index of $-3.28$. 
In order to confirm the lower limit of high-energy cut off in TeV region, 
we fitted the observed electron spectrum with an exponentially cut-off power law. 
With the fixed power-law index of $-3.28$, 
the lower limit (90~\% C.L.) of the exponential cut-off energy is $2.1$~TeV.

This observed electron spectrum in the energy region below 1~TeV 
is very similar to the electron spectrum by PAMELA \citep{adriani11a}, 
and shows agreement with the electron observations 
by Fermi-LAT \citep{ackermann10b}. 
Our electron spectrum observed with the emulsion chambers 
does not exhibit significant spectral excesses. 
The result of our electron spectrum compared with the ATIC electron spectrum 
\citep{chang08} is inconsistent, 
with a statistical significance level of $5$~\% 
(a reduced ${\chi}^{2}$ value of $1.834$ for d.o.f.$=11$), 
although it is acceptable with a statistical significance level of $1$~\%. 
Above 1~TeV region, 
the comparison of our result with the electron spectra by H.E.S.S. is consistent,
 with a reduced ${\chi}^{2}$ value of $1.121$ for d.o.f.$=7$, 
while H.E.S.S. data have large systematic errors \citep{aharonian08, aharonian09}.

\begin{figure}
\epsscale{1.00}
\plotone{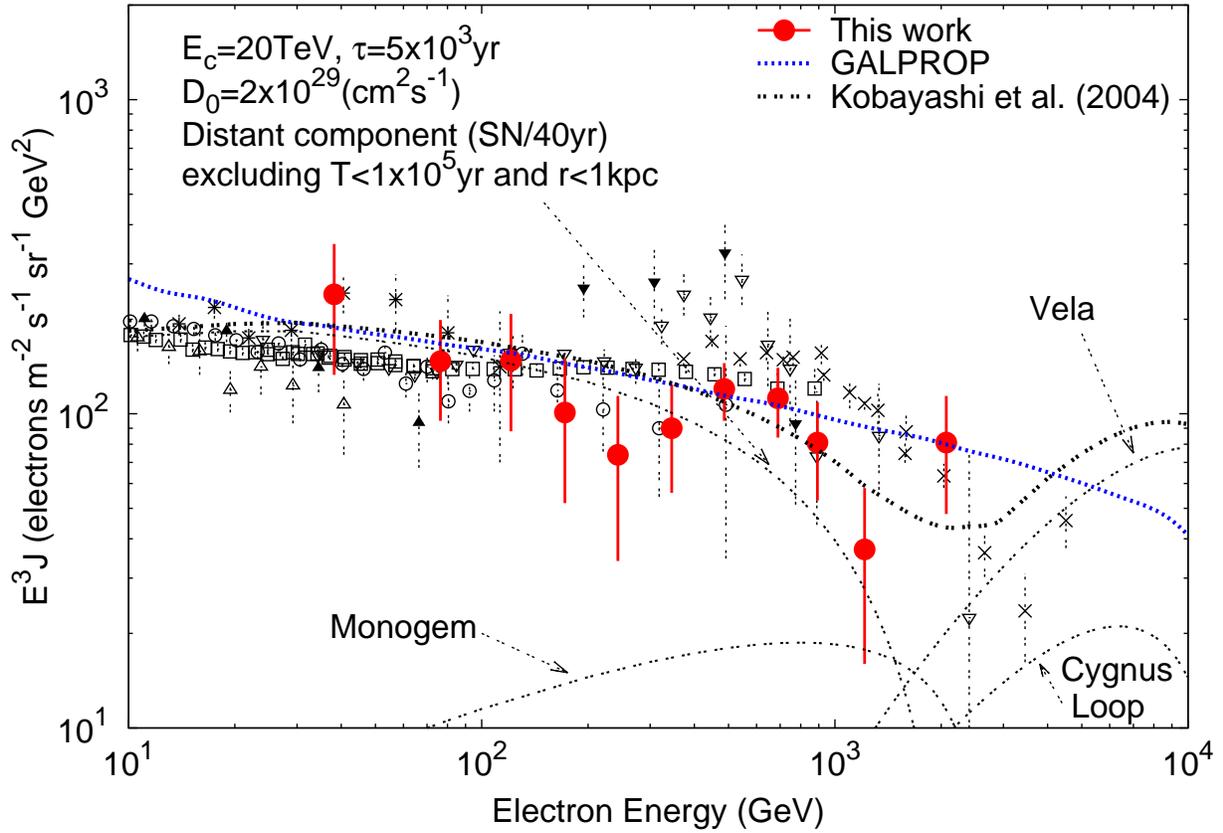}
\caption{
The observed electron spectrum with the emulsion chambers 
compared to model electron spectra, 
a GALPROP model with parameters of galdef\_50p\_599278 file 
and a SNRs model (a distant component $+$ nearby components) by \citet{kobayashi04}. 
See text for details. 
\label{fig:espec_models}}
\end{figure}

We calculated an electron spectrum by GALPROP code using a standard file, 
galdef\_50p\_599278 \citep{galprop10}. 
As shown in Fig.~\ref{fig:espec_models}, 
this spectrum is consistent with the electron spectrum 
observed with the emulsion chambers. 
We also compared the observed electron energy spectrum 
with an electron spectrum calculated by \citet{kobayashi04}, 
in which parameters are set as follows: 
the diffusion coefficient of $D_{0} = 2.0{\times}10^{29}$~cm$^2$~s$^{-1}$ at 1~TeV, 
the supernova rate of $1/40$~yr$^{-1}$ in the Galaxy, 
the electron output energy of $1{\times}10^{48}$~erg above 1~GeV, 
the 20~TeV cutoff energy of the electron injection spectrum, 
and the burst-like release at ${\tau}=5{\times}10^{3}$~yr after the explosion. 
The "distant component" in Fig.~\ref{fig:espec_models} 
indicates the contributions from continuously distributed distant SNRs 
with the distance larger than 1~kpc or the age older than $1{\times}10^{5}$~yr. 
As shown in Fig.~\ref{fig:espec_models}, 
our observed spectrum is consistent with the calculated spectra 
of the distant component $+$ nearby component by \citet{kobayashi04}, 
giving strong evidence for a non-zero flux in the TeV region, 
by the definite identification of electron events one by one. 
Figure~\ref{fig:espec_models} also presents 
that the electron energy spectrum observed with emulsion chambers 
has the significantly larger flux in the TeV region 
than that of the distant component. 
This suggests that nearby electron sources such as SNRs exist 
within the distance of $1$~kpc and the age of $1{\times}10^{5}$~yr. 

\begin{center}
\vspace*{0.5cm}
\framebox[3cm]{Figure \ref{fig:espec_models}}
\vspace*{0.5cm}
\end{center}

\section{Conclusions}

We have carried out cosmic-ray electron observations with the balloon-borne 
emulsion chambers since 1968. 
The emulsion chamber is an excellent imaging detector with a high position resolution 
of $1$~$\mu$m. 
This imaging capability enables the emulsion chamber to identify electrons 
with the high rejection power against gamma rays and protons, 
which is the highest proton rejection power of $1{\times}10^{5}$ 
among the existing cosmic-ray electron detectors, 
and to determine electron energies 
by using only the central part of the electro-magnetic shower. 
This leads for the emulsion chamber to have thin thickness, lower mass, 
and a wide field of view, 
compared to the other detectors. 
Hence, the emulsion chambers, even the comparatively lightweight instruments, 
successfully observed electrons above several 100~GeV in the late 1960s 
and electrons above 1~TeV in the 1980s. 
Further, these electron observations initiated discussions 
about the investigation of the propagation mechanisms in the Galaxy
and the identification of nearby cosmic-ray electron sources. 
Being recognized for the significance of the electron observations, 
the high-energy electron observations opened up 
and have been recently carried out by ATIC, Fermi-LAT, H.E.S.S., PAMELA, and so on. 
In order to identify nearby cosmic-ray electron sources and search for dark matter signals, 
there are also some ongoing and new experiments for the high-energy electron observations 
such as AMS-02 and CALET on the International Space Station (ISS) 
\citep{battiston08, torii11}. 
AMS-02 is ongoing to observe positrons and negative electrons up to 1~TeV. 
CALET is being developed to be installed on the ISS, 
preserving the excellent imaging characteristics of the emulsion chamber.

\vspace{1cm}

\acknowledgments

We sincerely thank the late Prof. J.J.Lord and the late Prof. T.Taira, 
who started up this experiment program, 
brought us to many successful balloon observations, 
and worked on the development of the experiments. 
We are also grateful to a number of collaborators 
who have carried out together the experiments since 1968. 
We thank the crews of the Sanriku Balloon Center (SBC) of ISAS/JAXA 
and the NASA Columbia Scientific Balloon Facility (CSBF) 
for their excellent and successful balloon flights. 
We acknowledge the staffs of 
Institute for Cosmic Ray Research (ICRR), University of Tokyo 
for their kind support to emulsion experiments. 
We also appreciate the excellent work and kind support of the staffs 
in Research Center for Electron Photon Science of Tohoku University 
and the H4 beam line of CERN-SPS. 
We are grateful to Y. Sato 
for his kind support at CERN in the beam tests.

\appendix

\section{A list of emulsion chambers}
\label{sec:ecc_list}

\begin{center}
\vspace*{0.5cm}
\framebox[3cm]{Table \ref{tab:ecc_list}}
\vspace*{0.5cm}
\end{center}

\section{Proton rejection power}
\label{sec:rejection}

The flux of cosmic-ray electrons is much smaller than 
that of cosmic-ray protons. 
The observed flux ratio of electrons against protons 
is around 1~\% for 10~GeV and 0.1~\% for 1 TeV.  
Therefore, 
in order to observe high-energy electrons above 1~TeV, 
the proton rejection power is required to be at least larger than $1{\times}10^4$. 
We estimate the proton rejection power with the emulsion chamber 
in the following.

\subsection{Mean free path for hadronic interactions}

While an electron produces the first  $e^{-}e^{+}$ pair within 2~r.l. 
with the probabilities of 99.9~\% for 100~GeV electrons 
and 99.5~\% for 1~TeV electrons 
with the threshold energy of 10~MeV for pair production, 
a proton interfaces hadronically with a mean free path (m.f.p.) 
of about 30~r.l. in lead. 
Hence, the probability for hadronic interactions of a proton within 2~r.l. 
in the emulsion chamber is $2/30$, 
so that the proton rejection power for the difference between interaction lengths 
of electrons and protons is about $15$.

\subsection{Energy shift of proton-induced shower}

Secondary neutral pions produced via hadronic interactions of a proton 
have in total about 30~\% of the parent proton energy. 
Therefore, giving the same shower energy with an electron, 
the energy of the parent proton should be three times larger than that of the electron. 
Since the integral cosmic-ray proton spectrum has a power-law form of $E^{-1.7}$, 
the effective flux of protons is reduced to be $0.3^{1.7}{\simeq}1/7$. 
Thus, the proton rejection power for the energy shift effect 
of the proton-induced shower is about 7.

\subsection{Shower starting point}

Electrons produce gamma rays via bremsstrahlung radiation, 
and then the gamma rays produce $e^{-}e^{+}$ pairs. 
Therefore, the almost electron-induced shower at the starting point 
in the emulsion chamber 
is composed of 3 tracks (one electron + one $e^{-}e^{+}$ pair) 
or 5 tracks (one electron + two $e^{-}e^{+}$ pairs). 
The mean spreading distance $r$ of a parent electron and an $e^{-}e^{+}$ pair 
by Coulomb scattering is approximately given by 
\begin{equation}
r \simeq \frac{1}{\sqrt{3}}\frac{E_{\rm s}}{E} (\frac{x}{X_{0}})^{1/2} L, 
\end{equation}
where $x$ is a traversing thickness of the electron in the material, 
$X_0$ is a radiation length of the material, 
$L$ is a path length of the electron, 
$E$ is the electron energy, 
and $E_{\rm s}$ is the scattering constant of ${\sim}20$~MeV. 
In the case of $x = X_{0}$ (i.e. $0.56$~cm for lead) and $L=1$~cm, 
the spreading distance $r$ is $2$~$\mu$m for the primary electron energy of 100~GeV 
and $0.2$~$\mu$m for that of 1~TeV, 
taking the electron energy $E$ to be a half of the incident electron energy 
because of the energy loss by bremsstrahlung radiation.

Among hadronic interactions of high-energy protons with nuclei, 
possible candidates to be the mimic electron events of protons are the following events. 
The forward produced charged secondary particles 
are narrowly collimated without the heavily ionizing tracks, 
which are recorded by low-energy recoil protons and nuclei, 
and the number of the charged secondary particles is just 1, 3 or 5, 
accompanied with one neutral pion. 
In addition, 
they have no multi-structures in their cascade showers. 
In fact, there are such phenomena 
that the momentum transfer to the target nuclei is relatively low 
and hence the number of the secondary particles is relatively small, 
what is called diffraction dissociation.

In order to study diffractive coherent production 
in hadronic interactions of protons with nuclei, 
the experiments with nuclear emulsions have been performed 
for 400 GeV and 800 GeV of proton beams by \citet{boos78} and \citet{abduzhamilov88}, 
respectively. 
They selected the events 
with the number of the secondary charged particles of 1, 3 or 5,  
with the relatively low momentum transfer to the target nuclei, 
and without the heavily ionizing tracks. 
According to their results, 
the fractions of these selected events to the total hadronic interaction events 
are 3~\% for 400~GeV protons and 2~\% for 800~GeV protons.

Given that the spreading angle of the forward produced secondary particles 
is $1{\times}10^{-3}$~rad, 
the radial distance between the secondary particles 
is $10$~$\mu$m with a path length of 1~cm. 
Since this distance is one order of magnitude larger than 
the typical distance between a parent electron and an electron-positron pair, 
the proton events with the spreading angle larger than $1{\times}10^{-3}$~rad 
are readily identifiable from the electron events. 
Adding the condition of the spreading angles less than $1{\times}10^{-3}$~rad, 
the residual events are less than 
$4$~\% of the measured diffraction dissociation events, 
that is, 
the fraction of the selected events to the total hadronic interaction events 
is less than 3~\%$\times$4~\% $=0.12$~\% for 400~GeV protons 
and 2~\%$\times$4~\%  $=0.08$~\% for 800~GeV protons.

For the electron-like events of protons, 
since there are also further conditions 
that the selected events should be accompanied with one neutral pion 
and have no multi-structures in the shower developments, 
the above fraction gives us just the upper limit. 
Hence, the proton rejection power with the shower starting point 
is estimated to be larger than $1/0.1$\% $= 1{\times}10^{3}$.

\subsection{Total proton rejection power} 

As described above, 
the proton rejection powers are $15$ for the difference 
between interaction lengths of electrons and protons, 
$7$ for the energy shift effect of the proton-induces shower, 
and $>1{\times}10^3$ for the shower starting point, respectively. 
Combined these independent proton rejection powers, 
the total proton rejection power is estimated to be larger than 
$15 {\times} 7 {\times} (1{\times}10^{3}) = 1{\times}10^{5}$. 
Thus, 
for the electron observations with the emulsion chambers, 
the proton contamination in the TeV region 
is estimated to be less than ${\sim}1$~\%.


\begin{table}[p]
\caption{\label{tab:baln_list} List of balloon flights}
\begin{center}
\begin{tabular}{lrrccr}
\hline\hline
Flight & Area & Time & Average Altitude & $S{\Omega}_{e}T$ $^*$ & 
Launch Site \\
       & (m$^2$) & (min) & (g~cm$^{-2}$) & (m$^2$ sr s) &  \\
\hline
1968   & 0.05 &  380 & 6.1 & $1.826\times10^3$ & Harunomachi, Japan \\
1969   & 0.05 &  267 & 7.1 & $1.283\times10^3$ & Harunomachi, Japan \\
1970   & 0.05 & 1136 & 6.1 & $5.460\times10^3$ & Sanriku, Japan \\
1973   & 0.20 &  833 & 8.2 & $1.934\times10^4$ & Sanriku, Japan \\
1976   & 0.40 & 1526 & 4.0 & $7.084\times10^4$ & Palestine, USA \\
1977   & 0.63 & 1760 & 4.5 & $1.2772\times10^5$ & Palestine, USA \\
1979   & 0.80 & 1680 & 4.9 & $1.5389\times10^5$ & Palestine, USA \\
1980   & 0.80 & 2029 & 7.8 & $1.8838\times10^5$ & Palestine, USA \\
1984   & 0.076 & 576  & 9.2 & $5.330\times10^3$ & Sanriku, Japan \\
1985   & 0.087 & 940  & 9.4 & $9.930\times10^3$ & Sanriku, Japan \\
1988   & 0.038 & 647  & 7.1 & $2.948\times10^3$ & Uchinoura, Japan \\
1996   & 0.20 & 2092 & 4.6 & $4.874\times10^4$ & Sanriku, Japan \\
1998   & 0.20 & 1178 & 5.6 & $2.729\times10^4$ & Sanriku, Japan \\
1999   & 0.20 & 891  & 5.6 & $2.005\times10^4$ & Sanriku, Japan \\
2001   & 0.20 & 1108 & 5.5 & $2.494\times10^4$ & Sanriku, Japan \\
\hline
\multicolumn{6}{l}{$^*$ Effective $S{\Omega}_{e}T$ 
for primary electron observations in the energy range above 1~TeV } \\
\multicolumn{6}{l}{within the zenith angle of 60$^{\circ}$. }

\end{tabular}
\end{center}
\end{table}

\clearpage

\begin{deluxetable}{cccccccc}
\tabletypesize{\footnotesize}
\rotate
\tablecaption{
The number of the observed electrons and the fluxes of primary cosmic-ray electrons 
\label{tab:elec_flux}}
\tablewidth{0pt}
\tablehead{
\colhead{Energy}     & \colhead{$\overline{E}$}   & \colhead{$S{\Omega}_{e}T$} & 
\colhead{$N_{ob}$}   & \colhead{$N_{sec}$}        & \colhead{$N_{pri}$}    & 
\colhead{Flux ($J$)} & \colhead{$E^{3}{\times}J$} \\
  (GeV)    & (GeV)           & (m$^{2}$ s sr) &         &           &           &  
(m$^{-2}$ s$^{-1}$ sr$^{-1}$ GeV$^{-1}$) & (GeV$^{2}$ m$^{-2}$ s$^{-1}$ sr$^{-1}$) \\
}
\startdata
30--50     & $3.82{\times}10^1$  & $69.8$      & 6        & 0         & 6         &
$(3.94{\pm}1.61){\times}10^{-3}$ & $220{\pm}90$  \\
60--100    & $7.64{\times}10^1$  & $682$       & 9        & 0         & 9         &
$(3.15{\pm}1.05){\times}10^{-4}$ & $141{\pm}47$  \\
100--150   & $1.21{\times}10^2$  & $1.679{\times}10^{3}$  & 8    & 1.00    & 7.00 &  
$(8.08{\pm}3.31){\times}10^{-5}$ & $143{\pm}59$  \\
150--200   & $1.72{\times}10^2$  & $5.613{\times}10^{3}$  & 7    & 1.43    & 5.57 &  
$(1.92{\pm}0.93){\times}10^{-5}$ & $98{\pm}47$  \\
200--300   & $2.43{\times}10^2$  & $9.718{\times}10^{3}$  & 7    & 1.96    & 5.04 &  
$(5.03{\pm}2.71){\times}10^{-6}$ & $72{\pm}39$  \\
300--400   & $3.45{\times}10^2$  & $4.8368{\times}10^{4}$ & 15   & 4.37   & 10.63 &  
$(2.14{\pm}0.80){\times}10^{-6}$ & $88{\pm}33$  \\
400--600   & $4.86{\times}10^2$  & $1.3374{\times}10^{5}$ & 35   & 6.44   & 28.56 &  
$(1.05{\pm}0.22){\times}10^{-6}$ & $120{\pm}25$  \\
600--800   & $6.90{\times}10^2$  & $3.2148{\times}10^{5}$ & 29   & 7.16   & 21.84 &  
$(3.35{\pm}0.85){\times}10^{-7}$ & $110{\pm}28$  \\
800--1000  & $8.92{\times}10^2$  & $5.9088{\times}10^{5}$ & 20   & 6.54   & 13.46 &  
$(1.13{\pm}0.39){\times}10^{-7}$ & $80{\pm}27$  \\
1000--1500 & $1.214{\times}10^3$ & $7.0795{\times}10^{5}$ & 15   & 7.73   & 7.27 &  
$(2.03{\pm}1.14){\times}10^{-8}$ & $36{\pm}20$  \\
1500--3000 & $2.068{\times}10^3$ & $7.0795{\times}10^{5}$ & 15   & 5.31   & 9.69 &  
$(9.04{\pm}3.74){\times}10^{-9}$ & $80{\pm}33$  \\
\enddata
\end{deluxetable}

\clearpage

\begin{deluxetable}{lcrlrrr}
\setcounter{table}{0}
\renewcommand*\thetable{A\arabic{table}}
\tabletypesize{\footnotesize}
\tablecaption{List of ECC \label{tab:ecc_list}}
\tablewidth{0pt}
\tablehead{
\colhead{ECC} & 
\colhead{X-ray film}   &             & 
\colhead{Area  (m$^2$)}    &   
\colhead{Edge effect}  & 
\colhead{${\theta}_{0}$ (deg)}  &
\colhead{$S{\Omega}_{e}T$ (m$^{2}$ sr s)} 
}
\startdata
1968   & (SN) & 
       &                                     &        &       &    \\
       &              &
$^{*}$ 30--50GeV:   & $19.49{\times}10^{-4}$ & 1.000  & 45  & $6.981{\times}10^{1}$  \\
       &             &
$^{*}$ 60--100GeV:  & $235.9{\times}10^{-4}$  & 0.807   & 45 & $6.819{\times}10^{2}$  \\
       &             & 
$^{*}$ 100--150GeV: & $235.9{\times}10^{-4}$  & 0.807  & 45  & $6.819{\times}10^{2}$  \\
       &             &  
$^{*}$150--200GeV: & $235.9{\times}10^{-4}$   & 0.807  & 45  & $6.819{\times}10^{2}$   \\
       &             &  
$^{*}$200--300GeV: & $467.7{\times}10^{-4}$   & 0.764   &  45 & $1.280{\times}10^{3}$  \\
       &             & 
$^{*}$ 300--400GeV: & $467.7{\times}10^{-4}$   & 0.764  & 45  & $1.280{\times}10^{3}$ \\
       &             & 
$^{*}$ 400--600GeV: & $467.7{\times}10^{-4}$   & 0.764  & 45  & $1.280{\times}10^{3}$ \\
       &             &  
600--800GeV: & $500.0{\times}10^{-4}$        & 0.680   & 60  & $1.826{\times}10^{3}$   \\
       &             & 
800--1000GeV: & $500.0{\times}10^{-4}$       & 0.680   & 60  & $1.826{\times}10^{3}$   \\
       &             & 
1000--1500GeV: & $500.0{\times}10^{-4}$      & 0.680   & 60  & $1.826{\times}10^{3}$   \\
       &             & 
1500--3000GeV: & $500.0{\times}10^{-4}$      & 0.680   & 60  & $1.826{\times}10^{3}$   \\
1969   & (SN) & 
       &                                     &        &        &    \\
       &        &     
30--50GeV:   & \nodata   & \nodata   &  \nodata  &  \nodata   \\
       &             &  
60--100GeV:  & \nodata   &\nodata    & \nodata   &  \nodata  \\
       &             & 
100--150GeV: & \nodata   & \nodata  & \nodata    &  \nodata  \\
       &             & 
150--200GeV: & \nodata   & \nodata   & \nodata   &  \nodata   \\
       &             &  
$^{*}$ 200--300GeV: & $318.3{\times}10^{-4}$   & 0.789   & 45 & $6.321{\times}10^{2}$ \\
       &             &  
$^{*}$ 300--400GeV: & $318.3{\times}10^{-4}$   & 0.789  & 45  & $6.321{\times}10^{2}$ \\
       &             &  
$^{*}$ 400--600GeV: & $318.3{\times}10^{-4}$   & 0.789  & 45  & $6.321{\times}10^{2}$ \\
       &             & 
600--800GeV: & $500.0{\times}10^{-4}$        & 0.680   & 60  & $1.283{\times}10^{3}$ \\
       &             &  
800--1000GeV: & $500.0{\times}10^{-4}$       & 0.680   & 60  & $1.283{\times}10^{3}$  \\
       &             &  
1000--1500GeV: & $500.0{\times}10^{-4}$      & 0.680   & 60  & $1.283{\times}10^{3}$ \\
       &             &  
1500--3000GeV: & $500.0{\times}10^{-4}$      & 0.680   & 60  & $1.283{\times}10^{3}$ \\
1970   & (SN) & 
       &                                     &        &       &     \\
       &        &   
30--50GeV:   & \nodata   & \nodata   &  \nodata  &  \nodata     \\
       &             & 
60--100GeV:  & \nodata   &\nodata    & \nodata    &  \nodata   \\
       &             & 
100--150GeV: & \nodata   & \nodata  & \nodata     &  \nodata   \\
       &             & 
150--200GeV: & \nodata   & \nodata   & \nodata     &  \nodata   \\
       &             &  
$^{*}$ 200--300GeV: & $115.9{\times}10^{-4}$   & 0.870   & 45  & $1.080{\times}10^{3}$ \\
       &             &  
$^{*}$ 300--400GeV: & $115.9{\times}10^{-4}$   & 0.870  & 45   & $1.080{\times}10^{3}$ \\
       &             &  
$^{*}$ 400--600GeV: & $115.9{\times}10^{-4}$   & 0.870  & 45   & $1.080{\times}10^{3}$ \\
       &             &  
600--800GeV: & $500.0{\times}10^{-4}$        & 0.680   & 60  & $5.460{\times}10^{3}$  \\
       &             &  
800--1000GeV: & $500.0{\times}10^{-4}$       & 0.680   & 60  & $5.460{\times}10^{3}$ \\
       &             &  
1000--1500GeV: & $500.0{\times}10^{-4}$      & 0.680   & 60  & $5.460{\times}10^{3}$  \\
       &             &  
1500--3000GeV: & $500.0{\times}10^{-4}$      & 0.680   & 60  & $5.460{\times}10^{3}$  \\
1973   & (SN) & 
       &                                     &        &       &     \\
       &        &   
30--50GeV:   & \nodata   & \nodata   &  \nodata   &  \nodata  \\
       &             &  
60--100GeV:  & \nodata   &\nodata    & \nodata    &  \nodata \\
       &             &  
$^{*}$ 100--150GeV: & $127.0{\times}10^{-4}$  & 1.000  & 45 & $9.972{\times}10^{2}$ \\
       &             &  
$^{*}$ 150--200GeV: & $127.0{\times}10^{-4}$  & 1.000   & 45  & $9.972{\times}10^{2}$ \\
       &             &  
$^{*}$ 200--300GeV: & $407.3{\times}10^{-4}$   & 0.873   & 45  & $2.792{\times}10^{3}$ \\
       &             &  
$^{*}$ 300--400GeV: & $407.3{\times}10^{-4}$   & 0.873  & 45   & $2.792{\times}10^{3}$ \\
       &             &  
$^{*}$ 400--600GeV: & $407.3{\times}10^{-4}$   & 0.873  & 45   & $2.792{\times}10^{3}$ \\
       &             &  
600--800GeV: & $0.20$        & 0.821   & 60   & $1.9335{\times}10^{4}$    \\
       &             &  
800--1000GeV: & $0.20$       & 0.821   & 60   & $1.9335{\times}10^{4}$    \\
       &             &  
1000--1500GeV: & $0.20$      & 0.821   & 60   & $1.9335{\times}10^{4}$   \\
       &             &  
1500--3000GeV: & $0.20$      & 0.821   & 60   & $1.9335{\times}10^{4}$   \\
1976   & (SN) & 
       &                                     &        &      &      \\
       &        &   
30--50GeV:   & \nodata   & \nodata   &  \nodata   &  \nodata  \\
       &             & 
60--100GeV:  & \nodata   &\nodata    & \nodata    &  \nodata \\
       &             & 
100--150GeV: & \nodata  & \nodata  & \nodata      &  \nodata \\
       &             & 
150--200GeV: & \nodata  & \nodata   & \nodata   &  \nodata   \\
       &             & 
200--300GeV: & \nodata   & \nodata  & \nodata   &  \nodata  \\
       &             & 
300--400GeV: & \nodata   & \nodata  & \nodata   &  \nodata  \\
       &             & 
400--600GeV: & \nodata  &  \nodata  &  \nodata  &  \nodata \\
       &             & 
600--800GeV: & $0.20$        & 0.821   & 60    & $3.5420{\times}10^{4}$    \\
       &             & 
800--1000GeV: & $0.40$       & 0.821   & 60    & $7.0841{\times}10^{4}$   \\
       &             & 
1000--1500GeV: & $0.40$      & 0.821   & 60    & $7.0841{\times}10^{4}$  \\
       &             & 
1500--3000GeV: & $0.40$      & 0.821   & 60    & $7.0841{\times}10^{4}$  \\
1977   &  (SN) & 
       &                                     &        &            \\
       &        & 
30--50GeV:   & \nodata   & \nodata   &  \nodata  &  \nodata   \\
       &             & 
60--100GeV:  & \nodata   &\nodata    & \nodata   &  \nodata  \\
       &             & 
100--150GeV: & \nodata  & \nodata  & \nodata     &  \nodata \\
       &             & 
150--200GeV: & \nodata  & \nodata   & \nodata    &  \nodata  \\
       &             & 
200--300GeV: & \nodata   & \nodata  & \nodata    &  \nodata  \\
       &             & 
300--400GeV: & \nodata   & \nodata  & \nodata    &  \nodata  \\
       &             & 
400--600GeV: & \nodata  &  \nodata  &  \nodata   &  \nodata  \\
       &             & 
600--800GeV:   & $0.04875$      & 0.810   & 60   & $9.824{\times}10^{3}$  \\
       &             & 
800--1000GeV:  & $0.24375$      & 0.810   & 60   & $4.9121{\times}10^{4}$    \\
       &             & 
1000--1500GeV: & $0.63375$      & 0.810   & 60   & $1.2772{\times}10^{5}$   \\
       &             & 
1500--3000GeV: & $0.63375$      & 0.810   & 60   & $1.2772{\times}10^{5}$  \\
1979   & (SN) & 
       &                                     &        &       &     \\
       &        & 
30--50GeV:   & \nodata   & \nodata   &  \nodata   &  \nodata  \\
       &             & 
60--100GeV:  & \nodata   &\nodata    & \nodata    &  \nodata \\
       &             & 
100--150GeV: & \nodata  & \nodata  & \nodata     &  \nodata \\
       &             & 
150--200GeV: & \nodata  & \nodata   & \nodata    &  \nodata  \\
       &             & 
200--300GeV: & \nodata   & \nodata  & \nodata    &  \nodata \\
       &             & 
300--400GeV: & \nodata   & \nodata  & \nodata    &  \nodata \\
       &             & 
400--600GeV: & \nodata  &  \nodata  &  \nodata   &  \nodata  \\
       &             & 
600--800GeV:   & $0.20$      & 0.810   & 60   & $3.8473{\times}10^{4}$    \\
       &             & 
800--1000GeV:  & $0.60$      & 0.810   & 60   & $1.1542{\times}10^{5}$    \\
       &             & 
1000--1500GeV: & $0.80$      & 0.810   & 60   & $1.5389{\times}10^{5}$  \\
       &             & 
1500--3000GeV: & $0.80$      & 0.810   & 60   & $1.5389{\times}10^{5}$  \\
1980   &  (SN) & 
       &                                     &        &       &     \\
       &        & 
30--50GeV:   & \nodata   & \nodata   &  \nodata   &  \nodata   \\
       &             & 
60--100GeV:  & \nodata   &\nodata    & \nodata    &  \nodata  \\
       &             & 
100--150GeV: & \nodata  & \nodata  & \nodata    &  \nodata   \\
       &             & 
150--200GeV: & \nodata  & \nodata   & \nodata   &  \nodata    \\
       &             & 
200--300GeV: & \nodata   & \nodata  & \nodata   &  \nodata  \\
       &             & 
300--400GeV: & \nodata   & \nodata  & \nodata   &  \nodata   \\
       &             & 
400--600GeV: & \nodata  &  \nodata  &  \nodata  &  \nodata   \\
       &             & 
600--800GeV:   & $0.30$      & 0.821   & 60   & $7.0644{\times}10^{4}$    \\
       &             & 
800--1000GeV:  & $0.80$      & 0.821   & 60   & $1.8838{\times}10^{5}$   \\
       &             & 
1000--1500GeV: & $0.80$      & 0.821   & 60   & $1.8838{\times}10^{5}$   \\
       &             & 
1500--3000GeV: & $0.80$      & 0.821   & 60   & $1.8838{\times}10^{5}$   \\
1984   & (SN), (G8) & 
       &                                     &        &      &      \\
       &        &    
30--50GeV:   & \nodata   & \nodata   &  \nodata   &  \nodata  \\
       &             & 
60--100GeV:  & \nodata   &\nodata    & \nodata    &  \nodata \\
       &             & 
100--150GeV: & \nodata  & \nodata  & \nodata    &  \nodata   \\
       &             & 
150--200GeV: & \nodata  & \nodata   & \nodata   &  \nodata   \\
       &             & 
200--300GeV: & \nodata   & \nodata  & \nodata   &  \nodata  \\
       &             & 
300--400GeV: & $381.0{\times}10^{-4}$   & 0.859  & 60   & $2.665{\times}10^{3}$  \\
       &             & 
400--600GeV: & $762.0{\times}10^{-4}$  &  0.859  & 60   & $5.330{\times}10^{3}$  \\
       &             & 
600--800GeV:   & $762.0{\times}10^{-4}$      & 0.859   & 60   & $5.330{\times}10^{3}$   \\
       &             & 
800--1000GeV:  & $762.0{\times}10^{-4}$      & 0.859   & 60   & $5.330{\times}10^{3}$    \\
       &             & 
1000--1500GeV: & $762.0{\times}10^{-4}$      & 0.859   & 60   & $5.330{\times}10^{3}$    \\
       &             & 
1500--3000GeV: & $762.0{\times}10^{-4}$      & 0.859   & 60   & $5.330{\times}10^{3}$    \\
1985   & (F), (G8) & 
       &                                     &        &        &    \\
       &        &  
30--50GeV:   & \nodata   & \nodata   &  \nodata   &  \nodata  \\
       &             & 
60--100GeV:  & \nodata   &\nodata    & \nodata    &  \nodata \\
       &             & 
100--150GeV: & \nodata  & \nodata  & \nodata     &  \nodata  \\
       &             & 
150--200GeV: & \nodata  & \nodata   & \nodata    &  \nodata   \\
       &             & 
200--300GeV: & \nodata   & \nodata  & \nodata    &  \nodata  \\
       &             & 
300--400GeV: & $762.0{\times}10^{-4}$   & 0.859  & 60  & $8.698{\times}10^{3}$   \\
       &             & 
400--600GeV: & $762.0{\times}10^{-4}$  &  0.859  & 60  & $8.698{\times}10^{3}$   \\
       &             & 
600--800GeV:   & $870.0{\times}10^{-4}$      & 0.859   & 60  & $9.930{\times}10^{3}$  \\
       &             & 
800--1000GeV:  & $870.0{\times}10^{-4}$      & 0.859   & 60  & $9.930{\times}10^{3}$  \\
       &             & 
1000--1500GeV: & $870.0{\times}10^{-4}$      & 0.859   & 60  & $9.930{\times}10^{3}$  \\
       &             & 
1500--3000GeV: & $870.0{\times}10^{-4}$      & 0.859   & 60  & $9.930{\times}10^{3}$  \\
1988   & (F), (G12), (GS), (G8) & 
       &                                     &        &      &      \\
       &        &  
30--50GeV:   & \nodata   & \nodata   &  \nodata   &  \nodata    \\
       &             & 
60--100GeV:  & \nodata   &\nodata    & \nodata    &  \nodata   \\
       &             & 
100--150GeV: & \nodata  & \nodata  & \nodata      &  \nodata  \\
       &             & 
150--200GeV: & $381.0{\times}10^{-4}$   & 0.846   & 60   & $2.948{\times}10^{3}$   \\
       &             & 
200--300GeV: & $381.0{\times}10^{-4}$   & 0.846  & 60    & $2.948{\times}10^{3}$ \\
       &             & 
300--400GeV: & $381.0{\times}10^{-4}$   & 0.846  & 60    & $2.948{\times}10^{3}$  \\
       &             & 
400--600GeV: & $381.0{\times}10^{-4}$  &  0.846  & 60    & $2.948{\times}10^{3}$  \\
       &             & 
600--800GeV:   & $381.0{\times}10^{-4}$     & 0.846   & 60  & $2.948{\times}10^{3}$  \\
       &             & 
800--1000GeV:  & $381.0{\times}10^{-4}$     & 0.846   & 60  & $2.948{\times}10^{3}$   \\
       &             & 
1000--1500GeV: & $381.0{\times}10^{-4}$     & 0.846   & 60   & $2.948{\times}10^{3}$   \\
       &             & 
1500--3000GeV: & $381.0{\times}10^{-4}$     & 0.846   & 60   & $2.948{\times}10^{3}$   \\
1996   & (F), (H8) & 
       &                                     &        &       &     \\
       &        & 
30--50GeV:   & \nodata   & \nodata   &  \nodata   &  \nodata   \\
       &             & 
60--100GeV:  & \nodata   &\nodata    & \nodata   &  \nodata   \\
       &             & 
100--150GeV: & \nodata  & \nodata  & \nodata    &  \nodata  \\
       &             & 
$^{*}$ 150--200GeV: & $50.0{\times}10^{-4}$   & 1.000   & 45  & $9.860{\times}10^{2}$   \\
       &             & 
$^{*}$ 200--300GeV: & $50.0{\times}10^{-4}$   & 1.000  & 45   & $9.860{\times}10^{2}$ \\
       &             & 
$^{*}$ 300--400GeV: & $50.0{\times}10^{-4}$   & 1.000  & 45   & $9.860{\times}10^{2}$  \\
       &             & 
400--600GeV:   & $0.20$     & 0.824   & 60   & $4.8735{\times}10^{4}$  \\
       &             & 
600--800GeV:   & $0.20$     & 0.824   & 60   & $4.8735{\times}10^{4}$   \\
       &             & 
800--1000GeV:  & $0.20$     & 0.824   & 60   & $4.8735{\times}10^{4}$   \\
       &             & 
1000--1500GeV: & $0.20$     & 0.824   & 60   & $4.8735{\times}10^{4}$  \\
       &             & 
1500--3000GeV: & $0.20$     & 0.824   & 60   & $4.8735{\times}10^{4}$   \\
1998   & (F), (H12), (H8) & 
       &                                     &        &       &     \\
       &        & 
30--50GeV:   & \nodata   & \nodata   &  \nodata   &  \nodata  \\
       &             & 
60--100GeV:  & \nodata   &\nodata    & \nodata    &  \nodata \\
       &             & 
100--150GeV: & \nodata  & \nodata  & \nodata     &  \nodata \\
       &             & 
150--200GeV: & \nodata   & \nodata   & \nodata   &  \nodata   \\
       &             & 
200--300GeV: & \nodata   & \nodata  & \nodata    &  \nodata  \\
       &             & 
300--400GeV:   & $0.20$     & 0.819   & 60   & $2.7287{\times}10^{4}$   \\
       &             & 
400--600GeV:   & $0.20$     & 0.819   & 60   & $2.7287{\times}10^{4}$  \\
       &             & 
600--800GeV:   & $0.20$     & 0.819   & 60   & $2.7287{\times}10^{4}$    \\
       &             & 
800--1000GeV:  & $0.20$     & 0.819   & 60   & $2.7287{\times}10^{4}$    \\
       &             & 
1000--1500GeV: & $0.20$     & 0.819   & 60   & $2.7287{\times}10^{4}$   \\
       &             & 
1500--3000GeV: & $0.20$     & 0.819   & 60   & $2.7287{\times}10^{4}$   \\
1999   & (F), (H16), (H12), (H8) & 
       &                                     &        &       &     \\
       &        & 
30--50GeV:   & \nodata   & \nodata   &  \nodata   &  \nodata   \\
       &             & 
60--100GeV:  & \nodata   &\nodata    & \nodata    &  \nodata  \\
       &             & 
100--150GeV: & \nodata  & \nodata  & \nodata     &  \nodata   \\
       &             & 
150--200GeV: & \nodata   & \nodata   & \nodata   &  \nodata    \\
       &             & 
200--300GeV: & \nodata   & \nodata  & \nodata    &  \nodata  \\
       &             & 
300--400GeV:   & \nodata    & \nodata  & \nodata  &  \nodata   \\
       &             & 
400--600GeV:   & $0.10$     &  0.796  & 60   & $1.0023{\times}10^{4}$  \\
       &             & 
600--800GeV:   & $0.20$     & 0.796   & 60   & $2.0046{\times}10^{4}$    \\
       &             & 
800--1000GeV:  & $0.20$     & 0.796   & 60   & $2.0046{\times}10^{4}$    \\
       &             & 
1000--1500GeV: & $0.20$     & 0.796   & 60   & $2.0046{\times}10^{4}$   \\
       &             & 
1500--3000GeV: & $0.20$     & 0.796   & 60   & $2.0046{\times}10^{4}$   \\
2001   & (F), (H16), (H8) & 
       &                                     &        &       &     \\
       &        & 
30--50GeV:   & \nodata   & \nodata   &  \nodata  &  \nodata    \\
       &             & 
60--100GeV:  & \nodata   &\nodata    & \nodata   &  \nodata   \\
       &             & 
100--150GeV: & \nodata  & \nodata  & \nodata     &  \nodata  \\
       &             & 
150--200GeV: & \nodata   & \nodata   & \nodata   &  \nodata    \\
       &             & 
200--300GeV: & \nodata   & \nodata  & \nodata    &  \nodata  \\
       &             & 
300--400GeV:   & \nodata    & \nodata  & \nodata &  \nodata     \\
       &             & 
400--600GeV:   & $0.20$     &  0.796  & 60   & $2.4937{\times}10^{4}$  \\
       &             & 
600--800GeV:   & $0.20$     & 0.796   & 60   & $2.4937{\times}10^{4}$    \\
       &             & 
800--1000GeV:  & $0.20$     & 0.796   & 60   & $2.4937{\times}10^{4}$    \\
       &             & 
1000--1500GeV: & $0.20$     & 0.796   & 60   & $2.4937{\times}10^{4}$   \\
       &             & 
1500--3000GeV: & $0.20$     & 0.796   & 60   & $2.4937{\times}10^{4}$  \\
\enddata
\tablecomments{$^{*}$ Microscope scanning. See text for details. }
\tablecomments{(SN) Sakura Type-N, (G8) G8-RX0, (F) Fuji \#200, 
(G12) G12-RXO, (H12) HR12-HA30, 
(GS) GS-RXO, (H8) HR8-HA30, (H16) HR16-HA30. }
\end{deluxetable}

\clearpage


\clearpage 

\clearpage

\clearpage 

\clearpage

\clearpage 

\clearpage 

\clearpage 

\clearpage

\begin{thebibliography}{}


\bibitem[Abdo et al.(2009)]{abdo09}
Adbo, A.A. et al. (Fermi-LAT Collaboration), 
2009, {Phys. Rev. Lett.} 102, 181101


\bibitem[Abduzhamilov et al.(1988)]{abduzhamilov88}
Abduzhamilov,A. et al. 
1988, {Mod. Phys. Lett. A} 3, 489



\bibitem[Ackermann et al.(2010a)]{ackermann10a}
Ackermann,M. et al. (Fermi-LAT Collaboration), 
2010, {\prd}, 82, 092003


\bibitem[Ackermann et al.(2010b)]{ackermann10b}
Ackermann,M. et al. (Fermi-LAT Collaboration), 
2010, {\prd}, 82, 092004


\bibitem[Adriani et al.(2009)]{adriani09}
Adriani,O. et al. 
2009, {Nature}, 458, 607


\bibitem[Adriani et al.(2011a)]{adriani11a}
Adriani,O. et al. 
2011a, {Phys. Rev. Lett.}, 106, 201101



\bibitem[Adriani et al.(2011b)]{adriani11b}
Adriani,O. et al. 
2011b, {Phys. Lett. B}, 703, 128


\bibitem[Agostinelli et al.(2003)]{agostinelli03}
Agostinelli,S. et al. 
2003, {Nucl. Instrum. Methods Phys. res. A}, 506, 250


\bibitem[Amako et al.(2006)]{amako06}
Amako,K. et al. 
2006, {IEEE Trans. Nucl. Sci.}, 53, 270


\bibitem[Aharonian et al.(2008)]{aharonian08}
 Aharonian,F. et al. (H.E.S.S. Collaboration), 
 2008, {Phys. Rev. Lett.}, 101, 261104


\bibitem[Aharonian et al.(2009)]{aharonian09}
 Aharonian,F. et al. (H.E.S.S. Collaboration), 
 2009, {\aap}, 508, 561


\bibitem[Aguilar et al.(2002)]{aguilar02}
 Aguilar,M. et al. 
 2002, {Physics Reports}, 366, 331


\bibitem[Amenomori et al.(2009)]{amenomori09}
Amenomori,M. et al. 2009, {\apj}, 692, 61


\bibitem[Baier \& Katkov(2005)]{baier05}
 Baier,V.N., \& Katkov,V.M., 
 2005, {Phys. Rep.}, 409, 261


\bibitem[Battiston et al.(2008)]{battiston08}
 Battiston,R. on behalf of the AMS-02 collaboration, 
 2008, {Nucl. Instr. Meth. A}, 588, 227


\bibitem[Beatty et al.(2004)]{beatty04}
 Beatty,J.J. et al. 
 2004, {Phys. ReV. Lett.}, 93, 241102


\bibitem[Boezio et al.(2000)]{boezio00}
 Boezio,M. et al. 
 2000, \apj, 532, 653


\bibitem[Boos et al.(1978)]{boos78}
 Boos,E.G. et al. 
 1978, {Nucl. Phys. B}, 137, 37


\bibitem[Chang et al.(2008)]{chang08}
 Chang,J. et al., 
 2008, {Nature}, 456, 362


\bibitem[Cheng et al.(2002)]{cheng02}
 Cheng,H.C., Feng,J.L., \& Matchev,K.T.
 2002, {Phys. Rev. Lett.}, 89, 211301


\bibitem[Daniel \& Stephens(1965)]{daniel65}
 Daniel,R.R., \& Stephens,S.A. 
 1965, {Phys. Rev. Lett.}, 15, 769


\bibitem[DuVernois et al.(2001)]{duvernois01}
 DuVernois,M.A. et al. 
 2001, \apj, 559, 296


\bibitem[Earl(1961)]{earl61}
 Earl,J.A.,  
 1961, {Phys. ReV. Lett.}, 6, 125


\bibitem[Golden et al.(1984)]{golden84}
 Golden,R.L. et al. 
 1984, \apj, 287, 622


\bibitem[Grimani et al.(2002)]{grimani02}
 Grimani,C. et al. 
 2002, {A\&A}, 392, 287


\bibitem[Haino et al.(2004)]{haino04}
 Haino,S. et al. 
 2004, {Phys. Lett. B}, 594, 35


\bibitem[Kamionkowski \& Turner(1991)]{kamionkowski91}
 Kamionkowski,M. \& Turner,M.S., 
 1991, {Phys. Rev. D}, 43, 1774

\bibitem[Kasahara(2012)]{kasahara12}
 Kasahara,K., 
 2012, {http://cosmos.n.kanagawa-u.ac.jp/}

\bibitem[Kobayashi et al.(1991)]{kobayashi91}
 Kobayashi,T. et al.  
 1991, {Nuovo Ciment A}, 104, 1753

\bibitem[Kobayashi et al.(2004)]{kobayashi04}
 Kobayashi,T., Komori,Y., Yoshida,K. \& Nishimura,J., 
 2004, {\apj}, {\bf 601}, 340.

\bibitem[Komori et al.(2012)]{komori12}
 Komori,Y., Kobayashi,T., Yoshida,K. \& Nishimura,J., 
 2012, {Astropart. Phys.}, in press 
(http://dx.doi.org/10.1016/j.astropartphys.2012.08.003)

\bibitem[Koyama et al.(1995)]{koyama95}
 Koyama,K. et al.  
 1995, \nat, 378, 255


\bibitem[Moskalenko \& Strong(2010)]{galprop10}
Moskalenko,I.V. \& Strong,A.W., 2010, 
{http://galprop.stanford.edu/}


\bibitem[Nishimura(1967)]{nishimura67}
 Nishimura,J., 
 1967, Handbuch der Physik, 46, II, 1, Springer


\bibitem[Nishimura et al.(1980)]{nishimura80}
 Nishimura,J. et al. 
 1980, \apj, 238, 394

\bibitem[Panov et al.(2011)]{panov11}
 Panov,A.D. et al. 2011, 
 {Astrophys. Space Sci. Trans.}, 7, 119


\bibitem[Shen (1970)]{shen70}
 Shen,C.S., 
 1970, \apj, 162, L181


\bibitem[Tang(1984)]{tang84}
 Tang,K.K., 1984, \apj, 278, 881


\bibitem[Torii et al.(2001)]{torii01}
 Torii,S. et al. 
 2001, \apj, 559, 973


\bibitem[Torii et al.(2008)]{torii08}
 Torii,S. et al. 
 2008, {arXiv:0809.0760v1 [astro-ph]}


\bibitem[Torii et al.(2011)]{torii11}
 Torii,S. on behalf of the CALET collaboration, 
 2011, {Nucl. Instr. Meth. A}, 630, 55


\bibitem[Yoshida et al.(2006)]{yoshida06}
 Yoshida,K., Ohmori,R., Kobayashi,T., Komori,Y., Sato,Y., and Nishimura,J.,  
 2006, {Phys. Rev. D}, 74, 083511


\bibitem[Yoshida et al.(2008)]{yoshida08}
 Yoshida,K., 
 2008, {Adv. Space Res.}, 42, 477


\end{thebibliography}
\end{document}